\documentstyle[aps,preprint,tighten,psfig]{revtex}

\begin{document}

\title{Hierarchical Structure of Azbel-Hofstader Problem:
 Strings and loose ends of Bethe Ansatz}

\author{A.G. Abanov, J.C. Talstra
\thanks{James Franck Institute of the University of Chicago,
5640 S.Ellis Avenue, Chicago, IL 60637, USA }
\and P.B. Wiegmann \thanks{James Franck Institute
and Enrico Fermi Institute of the University of Chicago, 5640 S.Ellis
Avenue, Chicago, IL 60637, USA and
Landau Institute for Theoretical Physics, Moscow, Russia}}

\maketitle

\begin{abstract}
We present  numerical evidence that solutions of the Bethe Ansatz
equations for a Bloch particle in an incommensurate magnetic field
(Azbel-Hofstadter or AH model), consist of complexes---{\it ``strings''}.
String solutions are well-known from integrable field theories. They become 
 asymptotically exact in the thermodynamic
limit.  The
string solutions for the AH model are exact in the incommensurate limit, where
the flux through the unit cell is an irrational number in units of the elementary flux
quantum.

We introduce the notion of the integral spectral flow and conjecture a
hierarchical tree for the problem. The hierarchical tree describes the topology
of the singular continuous
spectrum of the problem. We show that the string content of a state is
determined  uniquely
by the rate of the spectral flow (Hall conductance) along the tree. We identify
the Hall conductances with the set of Takahashi-Suzuki numbers (the set of
dimensions of the irreducible representations of $U_q(sl_2)$ with definite parity).

In this paper we consider the approximation of noninteracting strings. It
provides
the gap distribution function, the mean scaling dimension for the
bandwidths and
gives a very good approximation for some wave functions which even captures
their
multifractal properties.  However, it misses the multifractal character of the
spectrum.
\end{abstract}

\section{Introduction}

The problem of two-dimensional electrons in a periodic potential and a
uniform magnetic field (Azbel-Hofstadter problem \cite{Azbel64,Hofstadter76})
has a rich history and numerous applications. It
is equivalent (in the Landau gauge) to a one-dimensional quasiperiodic difference
equation:
\begin{eqnarray}
\label{harper}
   \psi_{n+1}+\psi_{n-1}
   +2\lambda\cos(k_{y}+2\pi n\eta)\psi_n=E\psi_n
\end{eqnarray}
with two competing incommensurate periods 1 and $1/\eta$. Here $\eta$ is the
flux of the magnetic field through the unit cell in units of the flux quantum
$h/e$. This equation is also known
as the Harper equation, the almost (or discrete) Mathieu equation, etc. It has been
applied to quasicrystals, localization/delocalization
transition\cite{AubryAndre80,Sinai87,LastJit94},  quantum Hall
effect\cite{ThoulessBook87} and even DNA  chains\cite{IguchiDNA97}.
For recent review of models of Hofstadter type see \cite{KreftSeiler96}
and references therein.

The spectrum of this equation is complex. If the flux $\eta$ is a rational
number $P/Q$, there is a common period $Q$ in the equation 
(\ref{harper}) and  the spectrum
consists of $Q$ bands. In the incommensurate limit, when  $\eta$ is an irrational
number\footnote{Although some properties of the spectrum depend on the
type of irrational number $\eta$, in this paper we consider
the so called {\it typically Diophantine} numbers. These numbers have a full
Lebesgue measure and thus suffice for almost all physical
applications.}  ($P\rightarrow
\infty,\,\,\,Q\rightarrow\infty$)  the spectrum becomes an infinite Cantor
set\footnote{closed,  nowhere dense set which has no isolated
points}\cite{Azbel64,BelSim82}   with total bandwidth (Lebesgue measure of the
spectrum) $4|\lambda -1|$
\cite{AubryAndre80,ThoulessBW83,AMS90}.  At $|\lambda| =1$ the spectrum
becomes a purely singular
continuous---uncountable but measure zero set of points\cite{Last94} 
(for review see \cite{Simon82,LastJit94}).  There
is  numerical evidence that in this case the spectrum and wave functions are
multifractal \cite{TangKohmoto83}.

The scaling description of multifractal sets, generated by a
quasiperiodic equation,  an important and challenging problem, is far
from being understood.  The reasons are twofold: a lack of language to
describe scaling properties of singular continuous sets and a lack of
methods to
attack incommensurate systems.

Recently, it has been shown that the Harper equation, belongs to the
class of integrable models of quantum field theory.  Despite being just
a one particle problem, it has been ``solved'' by the Bethe Ansatz
\cite{WiegmannZab94}.  This creates a perspective for an analytical solution
of the
problem and eventually for the employment of methods of conformal field theory
for finding scaling properties of multifractal sets.

Below we make a first step towards solving the Bethe Ansatz
equations for the most interesting critical case 
$|\lambda|=1$\footnote{The initial progress toward solving Bethe equations 
of \cite{WiegmannZab94} has been made in \cite{HKW94} where the explicit 
analytical solution for zero energy level as well as  some numerical 
results for midband levels are reported.}.  We show
that in the incommensurate limit roots of the  Bethe Ansatz equations are
grouped in complexes called {\em strings}, so that each state is a
composition of strings.  Strings are well known for standard integrable
models of quantum field theory, for instance XXZ Heisenberg chain
\cite{TakSuz72}.  They become exact only for a macroscopic system, where
the number of particles and the size of the system are sent to infinity.
The role of thermodynamic limit for the AH problem is
obviously the incommensurability---$P,Q \rightarrow \infty$, 
$P/Q\rightarrow \eta =\mbox{irrational}$. The common period $Q$ plays 
the role of the  size of the system.  Indeed, we present  numerical evidence that the
``string hypothesis'' remains valid---strings become exact in the
incommensurate
limit $Q\rightarrow\infty$.

A first important application of the string structure of solutions is a
detailed {\it hierarchical tree} of the spectrum, i.e.\  an
algorithm for generating this Cantor set spectrum\cite{tree}.  
The hierarchical tree
gives the topology of the set (Sec.\ \ref{strings}).  We show that the string
decomposition of a state is tied to the holonomy of the wave function,
i.e.\ the Hall conductance of the state, and therefore must be of an algebraic geometrical
nature.  The set of Hall conductances generates the spectral flow,
which in its turn describes the hierarchical tree  (Sec.\ \ref{HT}).

Since strings become exact in the incommensurate limit with an accuracy
${\cal O}(Q^{-2})$, at large but finite Q, the ``bare'' noninteracting
strings give a very good numerical approximation for some wave functions. 
This approximation even captures the multifractal properties of wave 
functions (Sec.\ \ref{NumEv}).  The noninteracting strings
give rise to a distribution function of gaps of the spectrum 
(Sec.\ \ref{gaps}).  The string
hypothesis also gives a mean field scaling exponent for the spectrum 
(bandwidth distribution), and
provides a framework for calculating anomalous dimensions as ``finite
size'' corrections to the bare values obtained at $Q\rightarrow\infty$ from
string solutions.  We will not compute them in this paper.

 We start by the
construction of the  hierarchical tree for the
spectral flow of a general quasiperiodic system (Sec.\ \ref{HT}). Then in
Sec.\ \ref{baesh} we review the Bethe Ansatz for the AH problem and formulate
the string solution.  In Sec.\ \ref{strings} we argue
that string solutions are  tied to the Hall conductances and generate
the hierarchical tree  described in Sec.\ \ref{HT}.  The string
hypothesis is used to compute the  gap and bandwidth distributions in
(Sec.\ \ref{gaps} and Sec.\ \ref{BWS}).  We
conclude with a list of results and ``loose ends''.   In the
Appendix (Sec.\ \ref{appendix}) we present a technically involved derivation of
Bethe Ansatz equations.

\section{Hierarchical-tree:
Spectral flow, Hall conductance and intermediate fractions.}
\label{HT}

The hierarchical tree is an adequate language to describe properties of
quasiperiodic systems and strange sets these systems generate.  In this
section we describe the hierarchical tree for the spectrum of the AH problem,
although a major part of the construction seems to be valid for more general
quasiperiodic linear systems.  Later in Sec.\ \ref{strings}  we show that this
tree determines the structure of the Bethe roots as well.

\begin{figure}
\centerline{\psfig{file=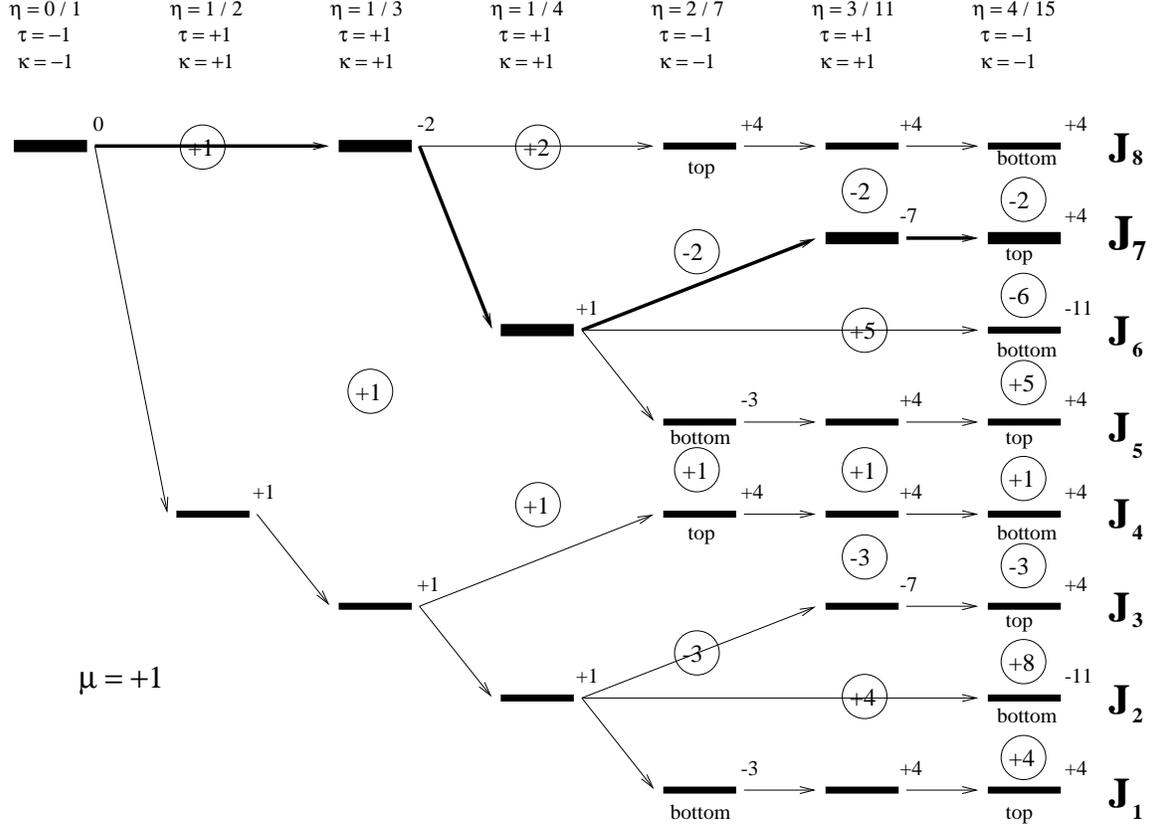,width=6.0in}}
\vspace{1.2cm}
\caption{The lower half of the hierarchical tree for the flux $\eta=4/15$
is shown. Nodes which lie on a vertical slice of the tree are bands
of a generation ordered with respect to their energies. 
The numbers in the figure are Hall conductances of bands and  gaps.
The Hall conductances in gaps  are shown as encircled numbers.
``Top'' and ``bottom'' are the edges of a band.
For all states shown $\mu=+1$.
Root patterns of  the highlighted branch $J_7$ of the tree 
 are shown in Fig.\ \protect\ref{fig.branch}.
}
\label{fig.htree}
\end{figure}

The hierarchical tree may be defined as follows.
Let us consider a sequence of rational
approximants $\eta^{(j)}=P^{(j)}/Q^{(j)}$ with increasing $Q^{(j)}$
to an irrational flux
$\eta$ so that:
$|\eta^{(j)}-\eta|<c(Q^{(j)})^{-2}$, where $c$ is
$j$-independent constant\footnote{This sequence always exists and
can be constructed e.g.\ from the so called Farey Series\cite{Vinogradov55}}. 
We refer to the spectrum of the 
Harper equation characterized by the  flux $\eta^{(j)}=P^{(j)}/Q^{(j)}$ 
as  the generation of the hierarchy. Let us
connect the $k$-th band of the generation $\eta^{(r)}$ (the daughter generation)
to a certain  band
$k'$ of some previous (parent) generation $\eta^{(r')}$, thus creating an
infinite graph with no loops. We call this graph a {\it hierarchical tree} if energies
$E_j({\cal J})$ belonging to any branch ${\cal J}$ of the tree form a
sequence converging to the point
$E({\cal J})$ of the spectrum in such a way that the sequence
$ (Q^{(j)})^{2-\epsilon_{\cal J}}|E_j({\cal J})-E({\cal J})|$ is bounded but
does not converge to zero.
The numbers $\epsilon_{\cal J}$ are anomalous
exponents\footnote{To avoid confusion
let us stress that the very existence of the hierarchical tree is a
(scaling) hypothesis. Moreover, the tree constructed in this paper is a
conjecture. There is no deeper reason for the hypothesis and the conjecture
other than physically plausible arguments and numerical evidence.}.
The example of hierarchical tree for AH problem 
is shown in Fig.\ \ref{fig.htree}.
To construct the hierarchical tree it is necessary to find the sequence of
generations and a rule to determine the parent band out of a given band of a
given generation.

Below we conjecture that the hierarchical tree is the spectral hierarchy---an
integral version of the spectral flow. Let us recall the notion of 
spectral flow and its rate---the Hall conductance. Consider the
spectrum of the problem (\ref{harper}) for a given flux
$\eta^{(r)}=P/Q$. It consists of Q bands separated by gaps. Let us choose some
gap
$k$ and change the flux by an infinitesimal $\delta\eta$. The spectrum
remains unchanged except
for some new levels that appear within a gap. The number $\delta N_k$ of
levels crossing the energy
$E$, lying inside the gap, is a spectral flow. Its rate
\begin{equation}
   \sigma_k = \frac{\delta N_{k}}{\delta\eta}
 \label{StredaF}
\end{equation}
is known to be the Hall conductance of the gap\cite{Streda}. This number
 is an integer and depends on the gap. The difference between the Hall
conductances of  neighboring gaps, i.e.\ the spectral flow into the $k$-th band
$\sigma(k)=
\sigma_{k}-\sigma_{k-1}$ is the Hall conductance of the band.

Let us now consider two close rational numbers  $\eta^{(j)}=P/Q$ and
$\eta^{(j-1)}=P'/Q'$ with
$Q'<Q$. The number of states in a band is $1/Q$ and $1/Q'$ respectively
in units of $N$, where $N$ is the total number of lattice sites. 
If there is a {\em band} $k$ of the problem with the flux $P/Q$, 
and we can find a flux $P'/Q'$
with $Q>Q',\, P\ge P'$ such that its Hall
conductance is a ratio between the difference of the number of states and the
fluxes
\begin{equation}
 \label{integrstreda}%
   \frac{1}{Q}-\frac{1}{Q'}=\sigma(k) (\frac{P}{Q}-\frac{P'}{Q'})
\end{equation}%
then we say that the band $k$ of the generation $P/Q$ has a ``parent''
band in the generation $P'/Q'$.
This formula may be viewed as an integrated Streda formula  (\ref{StredaF}).
It determines the flux $P'/Q'$ and by virtue of an iterative procedure,
generates a sequence of rational approximants (generations) $\eta^{(j)}$ to
the flux $\eta$.
The sequence  $\eta^{(j)}$ differs from the approximants obtained by
truncating the continued fraction expansion of
\begin{equation}
   \eta =
    \frac{1}{n_1+\frac{1}{n_2
    +\frac{1}{n_3+\ldots}}}
    \equiv [n_1,\,n_2,n_3,\ldots ].
  \label{parappr}
\end{equation}
Instead it consists of so called {\it intermediate fractions} approximants
(see below) \cite{Venkov70}.

We complete the integrated Streda formula by the {\it adiabatic} principle:
let us enumerate all states from
the bottom of the spectrum and characterize them by an integrated 
density of states (IDS)
$\nu=N_{\rm states}/Q$. Here $N_{\rm states}$ is  a total 
number of states with energy less than the energy of the given state. 
All states in the $k$-th band have the IDS
 $(k-1)/Q< \nu <k/Q$. The adiabatic principle assumes that a  middle 
 state\footnote{Middle state is the state for which the number of 
 states in the given band with lower energy is the same as number of 
 ones with higher energy.} in
 the band $k$ with IDS $(k-1/2)/Q$ has a parent in the
parent band $k'$ such that $(k'-1)/Q'<(k-1/2)/Q<k'/Q'$.
This gives
\begin{equation}
 \label{k}
   k' = \left[ \frac{Q'}{Q} (k-\frac{1}{2})\right]+1
\end{equation}
and together with integrated Streda formula determines the
hierarchical tree.
$\left[ x \right]$ denotes the integer part (floor function) of $x$.

To construct the tree, we need to know the Hall conductance of a band
$\sigma(k)$ and the Hall conductance of a gap
$\sigma_k$.  The Hall conductivity of the $k$-th gap at
generation $P/Q$
is known to be a solution of the Diophantine
equation\cite{TKNN82,DAZ85}
\begin{equation}
 \label{DE}
   P \sigma_k = k\; (\mbox{mod}\; Q)
\end{equation}
restricted to the range $-Q/2<\sigma_k\leq Q/2$.

To write the solution of this equation, let us represent the flux
$\eta^{(j)}=P/Q$ of the given generation as a
continued fraction (\ref{parappr}) and
 denote the convergents $P_i/Q_i=[n_1,\,n_2,\ldots
n_i],\,\,\,\,i=1,\ldots r;\;P_r/Q_r\equiv P/Q =\eta^{(j)}$.
Then the Hall conductance of a gap is
\begin{equation}
 \label{HC}
   {\sigma_k} = \frac{Q_r}{2}-Q_r\left\{(-1)^r  \frac{Q_{r-1}}{Q_r}k
   +\frac{1}{2}\right\},
\end{equation}
where $\{x\}$ is the fractional part of $x$.

In its turn the Hall conductivity  carried by
the $k$-th band ${\sigma}(k)=\sigma_{k} - \sigma_{k-1}$.
It is easy to see that the Hall conductivity of the band takes only two
values $(-1)^{r-1}Q_{r-1}$ and $(-1)^{r}(Q-Q_{r-1})$. It determines the
parent generation by virtue of (\ref{integrstreda})
$Q'=Q-|\sigma(k)|$:
\begin{equation}
\label{333}
\eta^{(j-1)}=\frac{P'}{Q'}=\left\{
\begin{array}{ll}%
\frac{P_r-P_{r-1}}{Q_r-Q_{r-1}},&\,\,\,\,
{\rm if}\,\,\,\,\sigma(k)=(-1)^{r-1}Q_{r-1}\\
\frac{P_{r-1}}{Q_{r-1}},&\,\,\,\,
{\rm if}\,\,\,\,\sigma(k)=(-1)^{r}(Q_{r}-Q_{r-1})
\end{array}\right.%
\end{equation}%

This formula looks better in terms of the continued
fraction
\begin{equation}\label{323}
\eta^{(j-1)}=\left\{
\begin{array}{ll}%
\left[n_1,n_2,\ldots,n_r-1\right],&\,\,\,\,
if\,\,\,\,\sigma(k)=(-1)^{r-1}Q_{r-1}\\
\left[n_1,n_2,\ldots,n_{r-1}\right],&\,\,\,\,
if\,\,\,\,\sigma(k)=(-1)^{r}(Q_{r}-Q_{r-1})
\end{array}\right.%
\end{equation}%
Thus the parent generation is obtained by either subtracting 1 from the last
quotient $n_r$ of the continued fraction or by truncating the fraction. The
sequence of generations produced by these iterations is known as the 
{\it intermediate fractions} . These numbers are defined as
numbers of the type $\frac{P_{i-2}+\mu P_{i-1}}{Q_{i-2}+\mu Q_{i-1}}$ with
$\mu = 0,1,2,\ldots,n_{i}$. The  intermediate
fractions list all the best rational approximants of the number
$\eta$\cite{ntnote}.
This is sufficient to construct the hierarchical tree.

Applying the same procedure to the parent generation we obtain another  set of values
of Hall conductances of bands. The entire set of Hall conductances generated by
the flux $P/Q$ are numbers less than $Q$
of the form
\begin{equation}
   \label{TZ}
   \{Q_{i-2}+mQ_{i-1}\;|\;m=1,\ldots,n_i,\;\;i=1,2,\ldots,r\}.
\end{equation}

This set  consists of numbers (denominators of intermediate fractions) known in
integrable models related to $U_q(SL_2)$ as {\em Takahashi-Suzuki} numbers
\cite{TakSuz72}.  They are allowed lengths of strings in solutions, say for
the XXZ model.  These numbers are also the set of possible dimensions of
irreducible highest weight  representations of $U_q(SL_2)$
with definite parity
\cite{MezNep90}.
We might have expected these numbers to play a role, considering that the intimate connection
between $U_q(Sl_2)$ and the AH-problem has already been outlined in previous
publications \cite{WiegmannZab94}.

Each path of the tree may also be characterized by a sequence of fractional 
numbers of
bands: $\nu^{(j)}_k=k/Q^{(j)}$, lying on the path and converging to a
given irrational fraction $\nu$. According to eq.(\ref{k}) the parent
fraction is determined by the daughter one as
$\nu^{(j-1)}=\frac{1}{Q^{(j-1)}}\left(\left[Q^{(j-1)}(\nu^{(j)}
-\frac{1}{2Q^{(j)}})\right]+1\right)$ with $Q^{(j-1)}=Q^{(j)}-|\sigma(k)|$. 
The sequence $\nu^{(j)}$ converges to the irrational $\nu$ faster than
$(Q^{(j)})^{-1}$,  $|\nu^{(j)}-\nu |<c (Q^{(j)})^{-1}$. In
terms of the fractions one may reformulate the scaling hypothesis as
$|E_j-E|<c|\nu^{(j)}-\nu |^{\alpha({\cal J})}$, defining the scaling exponent
$\alpha({\cal J})$.

To illustrate the construction of the hierarchical tree described above,
let us consider some particular flux, say, $\eta= P/Q=4/15=[3,1,3]$.  The
spectrum of AH problem for this flux consists of 15 bands and 14 gaps.
The ancestral generations of the tree are given by the sequence is
$\frac{0}{1}\,,\,\frac{1}{2}\,,\,\frac{1}{3}\,,\,
\frac{1}{4}\,,\,\frac{2}{7}\,,\frac{3}{11}\,,\,\frac{4}{15}$.  Notice, that the
truncated continuum fractions are different.  They are
$\frac{0}{1}\,,\,\frac{1}{3}\,,\,\frac{1}{4}\,,\,\frac{4}{15}$---just a subset
of generations.  The Hall conductances of
bands and gaps (ordered according to their energy) for each generation are:
\begin{eqnarray}
   \eta &=& \frac{4}{15}:\left\{
   \begin{array}{lll}
      \sigma(k) &=& 4,-11,4,4,4,-11,4,4,4,-11,4,4,4,-11,4\\
      \sigma_k &=& 4,-7,-3,1,5,-6,-2,2,6,-5,-1,3,7,4
   \end{array}
   \right.\nonumber\\
   \eta_6 &=& \frac{3}{11}:\left\{
   \begin{array}{lll}
      \sigma(k) &=& 4,-7,4,4,-7,4,-7,4,4,-7,4,\\
      \sigma_k &=& 4,-3,1,5,-2,2,-5,-1,3,-4
   \end{array}
   \right. \nonumber\\
   \eta_5 &=& \frac{2}{7}:\left\{
   \begin{array}{lll}
      \sigma(k) &=& -3,4,-3,4,-3,4,-3\nonumber\\
      \sigma_k &=& -3,1,-2,2,-1,3
   \end{array}
   \right.\\
   \eta_4 &=& \frac{1}{4}:\left\{
   \begin{array}{lll}
      \sigma(k) &=& 1,1,-3,1\\
      \sigma_k &=& 1,2,-1
   \end{array}
   \right. \nonumber\\
   \eta_3 &=& \frac{1}{3}:\left\{
   \begin{array}{lll}
      \sigma(k) &=& 1,-2,1\nonumber\\
      \sigma_k &=& 1,-1
   \end{array}
   \right.\nonumber\\
   \eta_2 &=& \frac{1}{2}:\left\{
   \begin{array}{lll}
      \sigma(k) &=& 1,-1\\
      \sigma_k &=& 1
   \end{array}
   \right.\nonumber
\end{eqnarray}
One should notice the asymmetry of $\sigma(k)$ for $\eta_4$. This asymmetry is 
related to the  degeneracy present for the fluxes with even denominators 
(two bands touch each other at energy $E=0$). The two possible 
choices $\sigma_2=\pm 2$
for the Hall conductance in the ``gap'' at $E=0$ (actually there is no gap) 
restore the symmetry of the tree with respect to $E\rightarrow -E$.
This tree, depicted in Fig.\ref{fig.htree} (half of the tree is shown) 
consists of 15 non-crossing 
branches $J_{i},\;i=1,2,\ldots 15$.
They connect bands, characterized by the flux and number of the band, 
with Hall conductances $\sigma(k)$ as listed below.  
The branches are listed in order of increasing energy. Only the lower 
half of the spectrum is presented
\begin{equation}
	\begin{array}{lll}
	J_1 &=& (\frac{1}{2},1,1; \frac{1}{3},1,1; \frac{1}{4},1,1; 
 	 	       \frac{2}{7},1,-3; \frac{3}{11},1,4; \frac{4}{15},1,4),	\\
           J_2 &=& (\frac{1}{2},1,1; \frac{1}{3},1,1; \frac{1}{4},1,1;
 	 	       \frac{4}{15},2,11),	\\
	J_3 &=& (\frac{1}{2},1,1; \frac{1}{3},1,1; \frac{1}{4},1,1; 
 	 	       \frac{3}{11},2,-7; \frac{4}{15},3,4),	\\
	J_4 &=& (\frac{1}{2},1,1; \frac{1}{3},1,1; 
 	 	       \frac{2}{7},2,4; \frac{3}{11},3,4; \frac{4}{15},4,4),	\\
	J_5 &=& (\frac{1}{3},2,-2; \frac{1}{4},2,1; 
 	 	       \frac{2}{7},3,-3; \frac{3}{11},4,4; \frac{4}{15},5,4),	\\
	J_6 &=& (\frac{1}{3},2,-2; \frac{1}{4},2,1; 
 	 	       \frac{4}{15},6,-11),	\\
	J_7 &=& (\frac{1}{3},2,-2; \frac{1}{4},2,1; 
 	 	       \frac{3}{11},5,-7; \frac{4}{15},7,4),	\\
	J_8 &=& (\frac{1}{3},2,-2; 
 	 	       \frac{2}{7},4,4; \frac{3}{11},6,4; \frac{4}{15},8,4).
	\end{array}
	\label{dcmp}
\end{equation}
Here in brackets we showed the flux, the number of the band and the 
Hall conductance of this band.

\section{Bethe Ansatz Equations and The String Hypothesis}
\label{baesh}

The Bethe Ansatz solution is available for any point of the Brillouin
z\^{o}ne $0\leq k_x<2\pi/Q,\;0\leq k_y<2\pi$  ($\eta=P/Q$) \cite{FadKash95},
but it looks especially simple in
the {\it rational} points of the Brillouin z\^{o}ne (the definition of 
rational points and corresponding values of $k_{x}$ and $k_{y}$ are 
given in Sec.\ \ref{appendix}).
They correspond to the centers and edges of bands.
The study of these points is sufficient for our purposes\footnote{We
also restrict ourselves to the isotropic (critical) case $\lambda=1$.}. 
Below we
sketch the results of the Bethe-Ansatz solution. The details can be found in
the Appendix and also in
Refs.\cite{WiegmannZab94,FadKash95,Zabrodin94}. The Harper eq.
(\ref{harper}) at rational points can be mapped  onto a functional equation
\begin{eqnarray}
   \mu \kappa z E\Psi(z)
   &=& iq (z+i\tau\kappa q^{-\frac{1}{2}})
          (z-i\kappa q^{-\frac{1}{2}}) \Psi(qz)
 \nonumber \\
   & & -iq^{-1} (z-i\tau\kappa q^{\frac{1}{2}})
                (z+i\kappa q^{\frac{1}{2}}) \Psi(q^{-1}z)
 \label{FullFuncEq}
\end{eqnarray}
where $q=e^{i\pi\eta}$ and $\tau,\,\kappa,\,\mu = \pm$ are labels of the
rational points (see Sec.\ \ref{appendix}).

The wave function of the original problem (\ref{harper}) can
be obtained as  a linear combination of $\psi_{n} = \Psi(- e^{-ip_-} q^n)$
where the momentum
$p_-$ belongs to a set of
rational points of the Brillouin z\^one . For
$P=\mbox{odd}$ the rational points correspond to the center of a band.
There are two distinct sets---one
for $\tau=+1$, another for $\tau=-1$.
For $P=\mbox{even}$ and $\tau=1$ a rational point also corresponds to a center
of the band. The rational points at
$\tau=-1$ and $P=\mbox{even}$,   give  bottom edges of odd
bands and top edges of even bands for $\mu (-1)^{P/2}=-1$, counted 
from the bottom
of the spectrum and  top  edges of odd bands and bottom edges of even bands
for  $\mu (-1)^{P/2}=+1$. This  allows one to find bandwidths and gaps. 
Let us take
$P=\mbox{even}$ and numerate the energy levels corresponding to rational 
points with $\tau=-1$ according to their order
starting from the
bottom of the spectrum $E_n,\;n=1,\ldots,Q $. Then the width  of the $n$-th
band ($W_n$) and the $n$-th gap ($D_n$) are:
\begin{eqnarray}
  \label{ww}%
   W_n=\mu (-1)^{\frac{P}{2}+n-1} (E_n+E_{Q+1-n}),
 \\
   D_n=\mu \left\{
   \begin{array}{ll}
      E_{n+1}-E_n\;\;\;\; &\mbox{if $P/2+n$ - odd} \\
      E_{Q-n+1}-E_{Q-n}\;\;\;\; &\mbox{if $P/2+n$ - even}
   \end{array}
   \right.  .
\end{eqnarray}%
 For $P$-odd and
$Q$-odd edges of bands and their widths may be obtained by the
flux reflection $q\rightarrow -q^{-1}$.

The integrability manifests itself in the fact that the solutions to the
difference equation (\ref{FullFuncEq}) are  polynomials
$$\Psi(z)=\prod_{i=0}^{Q-1} (z-z_i)$$
 with roots obeying Bethe equations:
\begin{equation}
   q^Q\prod_{j=1}^{Q-1}\frac{qz_i-z_j}{z_i-qz_j} =
   \frac{(z_i-i\tau\kappa q^{\frac{1}{2}})(z_i+i\kappa q^{\frac{1}{2}})}{
   (q^{\frac{1}{2}}z_i+i\tau\kappa)(q^{\frac{1}{2}}z_i-i\kappa)}.
 \label{BetheAnsatz2}
\end{equation}
These equations have $Q$ independent solutions. Each of them consists of
$Q-1$ complex
numbers $\{z_1,\ldots,z_{Q-1}\}$ and corresponds to an eigenstate of
the Harper equation at a rational point of the Brillouin z\^one
with the energy:
\begin{equation}
\label{energy0}
E=i\mu q^Q(q-q^{-1})
\frac{1}{2}\left[\kappa\sum_{i=1}^{Q-1}(z_i-q^{-Q}\tau z_{i}^{-1})
-i\frac{(1-q^{-Q})(1-\tau )}{q^{1/2}-q^{-1/2}}
\right].
\end{equation}

The Bethe equations (\ref{BetheAnsatz2}) are similar to the
equations for the XXZ chain with maximal allowed spin $S=\frac{Q-1}{2}$
and commensurate anisotropy $\cos (\frac{\pi}{2}\eta)$. The major
difference is that
it corresponds to a spin-problem on one or two sites
\cite{WiegmannZab94,YungBatchelor95}.

Below we give evidence that in an incommensurate limit all Bethe roots  are
grouped to strings. Each string of the ``spin'' $l$ is a complex of $2l+1$
roots
\begin{equation}
 \label{String2Def}
   \left\{ z_m^{(l)}=v_lx_l
   q_{l}^m,\,\,\,\,m=-l,-l+1.\ldots,l\right\}
\end{equation}
which have a common modulus $x_l>0$ (the center of the string), a parity $v_l=\pm
1$ and differ by multiples of $q_{l}$. We show that  $q_{l}$ is the $2l+1$-th
root of unity, ``closest'' to $q$ (Sec.\ \ref{strings}) and $x_l=1+{\cal
O}(Q^{-1})$.

Each state is characterized by a set of spins $\{l_k,l_{k-1},
\ldots\}$ and parities $\{v_{k},v_{k-1}, \ldots\}$ of the string content,
such that the total  number of roots
$\sum_{i=1}^k (2l_i+1)=Q-1$
\begin{equation}
\Psi^{\{l_k,v_k,\ldots\}}(z)\approx\prod_{m=-l_k}^{l_k}(z-x_lv_l
q_{l_k}^m)\Psi^{\{l_{k-1},v_{k-1},\ldots\}}(z).
\label{PsiAnsatz}
\end{equation}
Moreover, in the next section we argue that the wave function with 
a ``subtracted string''
$\Psi^{\{l_{k-1},v_{k-1},\ldots\}}(z)$ is approximately the wave 
function of the AH
problem of the parent generation, i.e.\ the problem with flux
$\eta'=\eta^{(j-1)}$ (\ref{323}) and the length of the ``subtracted''  
string is the Hall
conductance $l_k=|\sigma(k)|$. The formula (\ref{String2Def}) is
asymptotically exact with accuracy increasing with  the length of the string
as
$1/(2l_k+1)^2$.

Comparison with the XXZ chain may be useful.  In that case strings become
exact in the thermodynamical limit, i.e.\  in the limit where the size of
the chain and number of spins up go to infinity faster than anisotropy
angle $P/Q$ approaches an irrational number.  In that case spins (i.e.\
lengths) of the strings and their parities are drawn from a subset of
integers, called Takahashi-Suzuki numbers, while centers are continuously
distributed on the real axis.  Our case is different---the number of
sites on the equivalent chain is 2.  Also one does not expect to describe
strange sets by continuous distributions.  Nevertheless, the string
hypothesis remains robust.  Naturally the incommensurability plays the role
of ``thermodynamical limit''.  In the Sec.\ \ref{strings} we show that
allowed spins of strings remain to be the Takahashi-Suzuki numbers, however
the number of strings with a given spin is finite and their centers
approach 1 with increasing length of the string.

Strings play a bigger role than just being solutions of the Bethe
equations.  The spins of strings of which the state is composed give the
topological characteristics of the state, related to the holonomy of the
wave function in the Brillouin z\^one (the Hall conductance).  They
eventually describe the topology of the spectrum.

\section{Strings and Hierarchy Tree}
\label{strings}

In this section we present heuristic arguments for the string hypothesis
and support them with numerics in Sec.\ \ref{NumEv}. The arguments are based
on the analysis of singularities of the Bethe equations
(\ref{BetheAnsatz2}). It turns out that the string solutions follow the
hierarchical tree of Sec.\ \ref{HT}.

Let us assume that a solution of the Bethe equations (\ref{BetheAnsatz2})
contains a long string of spin $l\gg 1$ (\ref{String2Def}). Let us
substitute this string into (\ref{BetheAnsatz2}) and drop all terms, which are not singular as
$q_l\rightarrow q$. Equations for the first and the last root of the
string $v_lx_lq_l^{-l}$ and $v_lx_lq_l^l$ give
\begin{eqnarray}
   \kappa x_lv_lq_l^{-l}(q-q_l)(\ldots)\sim
   (x_lv_lq_l^{-l}-i\kappa
q^{\frac{1}{2}}\tau)(x_lv_lq_l^{-l}+i\kappa q^{\frac{1}{2}})
  \label{111}   \\
   \kappa x_lv_lqq_l^{l-1}(q_l-q)(\ldots)\sim
   (q^{\frac{1}{2}}x_lv_lq_l^l+i\kappa\tau)(q^{\frac{1}{2}}x_lv_lq_l^l-i\kappa).
  \label{112}
\end{eqnarray}
where by $(\ldots)$ we denote the terms of the order one at $q_l\rightarrow q$.
Compatibility of these equations requires
\begin{equation}
 \label{z}
   |q_l^{2l}q\pm 1|={\cal O}(|q-q_l|).
\end{equation}
If $q_l=e^{i\pi P''/Q''}$ is sufficiently close to $q=e^{i\pi\eta}$
i.e.\ $|\eta-P''/Q''|\sim 1/Q''^2$, Eq. (\ref{z}) may be
satisfied only if
\begin{equation}
 \label{stringq}
   q_l^{2l+1}=\mp 1
\end{equation}
where $\mp$ corresponds to the sign in (\ref{z}). This gives $Q''=2l+1$.
We will refer to the strings
with $q_l^{2l+1}= +1$ as to ``closed'' strings as opposed to ``open'' ones
with
$q_l^{2l+1}= -1$.

For open strings the equations (\ref{111},\ref{112})
give an approximate value of the center $x_l\approx 1$ and
two possible values of parity:
\begin{eqnarray}
 \label{v}
   v_l  &=&  -iq_l^{l+1/2}\kappa=(-1)^{\left [\eta l\right]}\kappa , \\
 \label{vprime}
   v_l  &=&  +iq_l^{l+1/2}\tau\kappa=-(-1)^{\left [\eta l\right]}\tau\kappa .
\end{eqnarray}
The  symmetries of the
Bethe equations require a center to be a real positive number. 
In case $x_l\neq 1$,
there is always another string with an inverted center $x_l^{-1}$.

If $q_l^{2l+1}= 1$ the leading singular terms
(\ref{111},\ref{112}) of the Bethe equations vanish. The center  and the parity
of the closed string  are determined by the next leading singularities. We do
not study these here.  Empirically we still find that centers of closed
strings are
close to $1$ as for open strings.

Now consider the Bethe equation for a root which does not
belong to the string.
Then, assuming $q_l$ to be very close to $q$ after a set of
cancelations, we obtain again the Bethe equations for
$Q-2l-2$ remaining roots.  If $q_l^{2l+1}=1$ the equations retain their form.
In case
$q_l^{2l+1}=-1$ the Bethe equations for the point $(\tau,\kappa,\mu)$ are 
transformed into ones for the point $(-\tau,-\kappa,\mu)$ for a string with
the parity (\ref{v}) and to the point  $(-\tau,\kappa,-\mu)$ for the 
parity (\ref{vprime}).

In other words, while subtracting the string from a wave function of a generation where
rational points correspond to edges of  bands ($P$ is even) we obtain the Bethe
equation for the generation where rational points are the centers of bands
($P'$
is odd). However, subtracting a string from the generation where rational
points
are centers ($P$ is odd) one may get $P'$ both even and odd, depending on the
sign in eq.(\ref{stringq}).

Now comes the crucial step: we have to determine what values $2l+1$, the length
of the string and its parity can take.
The  new approximate Bethe equations on the remaining $Q-2l-2$ roots
would be consistent if one replaces $q$ by the closest $q'$  such  that
$(q')^{Q-2l-1}=\pm 1$. This requirement and also the requirement that
$q_l$ is also the closest to $q$, leaves us with two possibilities
\begin{equation}
 \label{444}
   \begin{array}{l}
    q_l= e^{i\pi\frac{P_{r-1}}{Q_{r-1}}} \\
    q'= e^{i\pi\frac{P-P_{r-1}}{Q-Q_{r-1}}}
   \end{array}
   \;\;\;\mbox{or}\;\;\;
   \begin{array}{l}
    q_l= e^{i\pi\frac{P-P_{r-1}}{Q-Q_{r-1}}} \\
    q'= e^{i\pi\frac{P_{r-1}}{Q_{r-1}}}
   \end{array}.
\end{equation}
This means that the states of generation $P/Q$ have the largest string of
the length
\begin{equation}
\label{555}
2l+1= \{Q-Q_{r-1}\;\;\;\mbox{or}\;\;\;Q_{r-1}\}
\end{equation}
The parent generation of a state may have a flux:
\begin{equation}
 \label{666}
   \eta^{(r-1)}=\frac{P_{r-1}}{Q_{r-1}}\;\;\;\mbox{or}\;\;\;
   \frac{P-P_{r-1}}{Q-Q_{r-1}},
\end{equation}
respectively. Thus we see that  the
lengths of two possible highest strings are the same as two possible values of
Hall conductance of  bands (\ref{333}). We conjecture (and checked this
numerically) that the length of the highest string is equal to the Hall
conductance of the band.

Now let us discuss the parities of open strings. If $\tau=-1$ the 
both formulae (\ref{v},\ref{vprime}) give the same result 
$v=(-1)^{[\eta l]}\kappa$. However for $\tau=+1$ (center of bands) there is 
a choice between two possible values of parity. To make this choice we
must further specify the hierarchical tree and the spectral flow. 
 Let us consider for example the band $k$ 
which belongs to the 
generation $\eta=P/Q$ with $P$-odd and $\tau=+1$. Then the corresponding 
parent band $k'$ belongs
to the generation $\eta'=P'/Q'$ with $P'$-even (we consider the case when the 
largest string is open) and $\tau'=-\tau=-1$. There are two different
energy levels corresponding to the rational points in Brillouin z\^one
which belong to the parent band $k'$. These are top and bottom of this 
band. We will call the state corresponding to the top (or bottom) of the 
band $k'$ as {\em parent state} of the state in the middle of the $k$-th
band if the Hall conductance in the gap adjacent to the top (bottom) of the 
band $k'$ is the same as the one in the gap adjacent to the top (bottom)
of the band $k$ of the generation $\eta$. If one goes 
from the generation $\eta'$ to generation $\eta$ in AH problem only new
gaps open in the spectrum while the old ones 
present in parent generation stay and have the same Hall conductance in
the daughter generation. Therefore among two states at the top and the 
bottom of the parent band we call the parent state the one which is not
splitted of from the band by new gap while evolving from the generation
$\eta'$ to the generation $\eta$. The both states can be parent states 
if the band $k'$ is not split by a new gap.

The edge of the band $k'$ corresponding to the rational point characterized
by $\kappa'$ and $\mu'$ is a top of the band if $(-1)^{P'/2+k'}\mu'=-1$ and it 
the bottom if $(-1)^{P'/2+k'}\mu'=+1$. Using the rules of transformation
of $\kappa$ and $\mu$ while subtracting open string from the state in the 
middle of the band $k$ one can easily obtain that the right choice of 
parity is $v_l=\mu\kappa (-1)^{P'/2+k'+[\eta l]}$ if the parent state is the
bottom of the band $k'$ and $v_l=-\mu\kappa (-1)^{P'/2+k'+[\eta l]}$ if 
it is the top. Finally, the $k$-th gap at the generation $\eta$ is an old
one (present at the generation $\eta'$) if corresponding Hall conductance
$|\sigma_k|\le Q'/2$. and we obtain:
\begin{equation}
 \label{vppp}
   v_l=\mu\kappa (-1)^{P'/2+k'+[\eta l]} \left\{
   \begin{array}{lll}
     +1 & \;\;\;\;\mbox{if}\;\;\;\; & |\sigma_{k-1}|\le Q'/2 \\
     -1 & \;\;\;\;\mbox{if}\;\;\;\; & |\sigma_{k}|\le Q'/2 
   \end{array}
   \right.  .
\end{equation}
If both of formulae in (\ref{vppp}) work one can choose parity to be $v=\pm 1$.
We notice than in this case the string content of the state in the middle 
of the band $k$  must include the string of the largest length at least twice
with both possible parities. 

Summing up we obtain the {\it iterative procedure} 
of constructing the string decomposition:
\begin{itemize}
\item
Starting from a band of the generation $\eta^{(r)} =P_r/Q_r$ and a
point of the Brillouin z\^one $\tau,\kappa,\mu$, we determine the string with
highest spin (\ref{555}). Its parity is given by (\ref{v}).
Subtracting this string we obtain the Bethe equation for a generation
$\eta^{(r-1)}$ (\ref{666}) for a point of the Brillouin z\^one
$\tau'=q_l^{2l+1}\tau, \kappa'=\pm \kappa$ and so on.  Continuing this
procedure to the end of the tree we obtain an unambiguous string decomposition
of each level.
\end{itemize}

To illustrate the iterative hierarchical procedure, let us denote by
$\left[\frac{P}{Q},\tau,\kappa,\mu \right]$ the set of  roots ($Q-1$ complex
numbers) for flux $P/Q$ and rational points of a particular band labeled by
$\tau$,$\kappa$ and $\mu$. Then the iterative procedure 
can be symbolically written as
\begin{equation}
   \left[\frac{P}{Q},\tau,\kappa,\mu\right]  \rightarrow
   \{v_l\cdot({\bf 2l+1})\}\oplus
   \left[\frac{P'}{Q'},q_l^{2l+1}\tau,\pm \kappa,\pm \mu\right]
 \label{hier1}
\end{equation}
Here the symbol $\{v_l\cdot({\bf 2l+1})\}$ denotes the string of the length
$2l+1$  with the parity $v_l$, $Q'=Q-2l+1,\;P'=P-P''$ and $P''$ is the nearest
integer to $(2l+1)\eta$. If the string is open, i.e.\ $q_{l}^{2l+1}=-1$ the
numerators $P$ and $P'$ have opposite parity (if $P$ is odd, $P'$ is even and
vice versa), while $P$ and $P'$ have the same parity if the string is closed.
The symbolic expression (\ref{hier1}) means that Bethe roots corresponding 
to the state $\left[\frac{P}{Q},\tau,\kappa,\mu\right]$ are given by the roots of 
the  state
$\left[\frac{P^{'}}{Q^{'}},q_{l}^{2l+1}\tau,\pm\kappa,\pm\mu\right]$ plus the
roots of the string of the length $2l+1$ with parity $v_l$.

The  iterative procedure of the string's decomposition is identical to
the hierarchical tree based on the spectral flow (Sec.\ \ref{HT}). Once
the length of the highest string in a given band is the Hall
conductance of the band, the string length is the rate of the spectral flow from
the parent bands. Subtracting this string we arrive at the parent band
of the generation given by the second term of the sequence of
intermediate fractions (\ref{323})).  Performing this procedure recursively, we obtain the entire set of
possible lengths are the celebrated Takahashi-Suzuki numbers (\ref{TZ}))
known from the  $U_q(Sl_2)$ integrable models.

Since lengths of strings are Hall conductances, the string content of a state
can be read off from the hierarchical tree. The lengths of these strings are the Hall
conductances of the bands on a branch (connecting these bands) that leads to the state. The string
decomposition of the 15 bands of the example of the Sec.\ \ref{HT}
$P/Q=4/15$ (\ref{dcmp}) is, as shown in  Fig.\ \ref{fig.htree} (only the lower
half of the tree is shown):
\begin{equation}
\begin{array}{llllll}
   J_1 &=& (-1,-1,-1,-3,4,-4), &   J_9    &=& (\pm 2,1,-7,-4),       \\
   J_2 &=& (-1,-1,-1,11),    &   J_{10} &=& (\pm 2,1,11),        \\
   J_3 &=& (-1,-1,-1,7,-4),   &   J_{11} &=& (\pm 2,1,3,-4,-4),     \\
   J_4 &=& (-1,-1,-4,4,-4),    &   J_{12} &=& (1,1,4,-4,-4),    \\
   J_5 &=& (\pm 2,-1-3,4,-4),   &   J_{13} &=& (1,1,1,-7,-4),   \\
   J_6 &=& (\pm 2,-1,11),     &   J_{14} &=& (1,1,1,11),    \\
   J_7 &=& (\pm 2,-1,7,-4),    &   J_{15} &=& (1,1,1,3,-4,-4), \\
   J_8 &=& (\pm 2,-4,4,-4),    &          & &
\end{array}
\label{dcmp1}
\end{equation}%
where in brackets we showed the lengths of strings and their parities 
as signs in front of lengths.

\begin{figure}
\centerline{\psfig{file=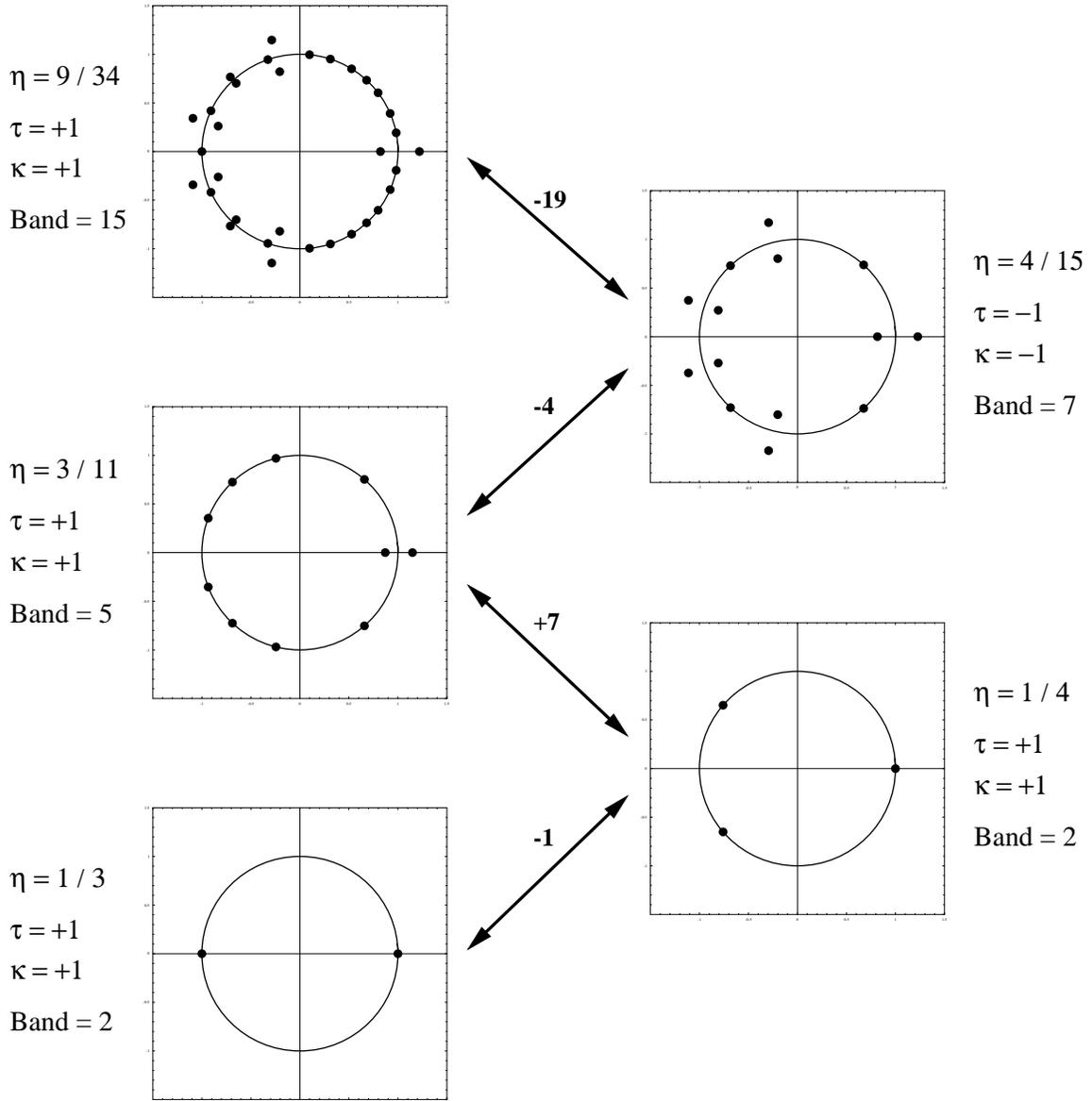,width=6.0in}}
\vspace{0.2cm}
\caption{The root patterns along one branch ($J_7$) of the hierarchical tree
of Fig.\ \protect\ref{fig.htree}. Subsequent root patterns along the tree
differ by a string whose length is equal to the Hall conductances of
the ``younger''-one of the two bands. The parities of these strings are in agreement with
(\protect\ref{v}) and are shown as signs in front of string lengths.
The band numbers are given as counted from the bottom of the spectrum.
For all states shown $\mu=+1$. 
}
\label{fig.branch}
\end{figure}

Fig.\ \ref{fig.branch} shows the root patterns of the 7-th branch 
(counted from the bottom of the spectrum)
$J_7$  of the tree from Fig.\ \ref{fig.htree}. We added also the root
pattern of
the 15-th level for the flux
$9/34$ which belongs to the same sequence. One can notice that the subsequent
 patterns of roots differ only by the string determined by
hierarchy procedure. Although the topology of the hierarchical tree seems to be
exact, its quantitative description in terms of strings is approximate. The
true
values of roots deviate from their bare value $x_l=1$ (see Fig.\
\ref{fig.branch}).
Eq.(\ref{String2Def}) is approximate and valid only
for long strings.

\section{Gap distribution and bandwidths}
\label{gapsbands}

\subsection{Gaps}
\label{gaps}

A direct application of the string hypothesis is the gap distribution
$\rho(D)$, i.e.\ the number of gaps with magnitude between $D$ and $D+dD$. This problem is
essentially reduced to the scaling of the smallest gaps. Indeed, while moving
along the tree new gaps appear, while older gaps become wider. There is no
overlap between bands except at $E=0$ for even denominators of the
flux (they touch)---all gaps
are open \cite{Mouche89}.

Thus while moving from a generation $P'/Q'$ to generation
$P/Q$, there  will be $Q-Q'$ new gaps. For typically Diophantine flux
denominator grows fast with the generation number and $Q-Q'\sim Q$. This
means that the integral
$\int^{D_{\rm max}}_{D_{\rm min}}\rho(D)dD=Q$  is
saturated by the lower limit. If $\rho\sim D^{-\gamma}$, then
$D^{-\gamma+1}_{\rm min}\sim Q$.

To find the size of the minimal gap,
let us consider $P$-even and  $\tau=-1$. Then the Bethe       
equations give edges of gaps.   Let us consider an arbitrary gap in
this generation and trace
the genealogy of lower and upper edges of the gap along two paths $J_-$ and
$J_+$ of the tree, until we find the nearest common ancestor at the
generation with denominator
$Q'$. The minimal gap would correspond to the shortest paths $J_{\pm}$, 
i.e.\ when 
$Q'$ is the denominator of the generation  parent to $P/Q$.
 The string decompositions of the two edges
have a
common part consisting of strings of length smaller than $Q'$ (they
correspond to the common path from the origin of the tree to the parent
generation $Q'$) and {\em different strings} with length bigger than $Q'$
belonging to the paths $J_{\pm}$.  Then according to (\ref{energy0}), the
width of the gap is the difference between
energies of  strings with length greater than $Q'$ along the path $J_-$
and the energies of strings along $J_+$. The contribution to the energy $\varepsilon_l$
from a string with a spin $l$ which connects the center of
the band ($\tau=+1$)  with the edge of the band ($\tau=-1$) is
\begin{equation}
   \varepsilon_l= i(q-q^{-1})
   2\left(\sum_{k=-l}^{l}x_lv_lq_l^{k}-\frac{2i}{q^{1/2}-q^{-1/2}}\right)
   \approx \frac{4\pi}{Q(2l+1)} \frac{\cos^2(\frac{\pi}{2}\frac{P}{Q})}
   {\sin(\frac{\pi}{2}\frac{P}{Q})}\sim \frac{1}{Q^2} ,
 \label{energy03}
\end{equation}
where we used the fact that for a typical Diophantine flux $\eta$ the length of the 
largest string $2l+1$ scales as $Q$ along the sequence of rational approximants. 
This  gives us an estimate for the gap $E_l\sim  1/Q^2$.
The difference of energies of strings is of  the same order. This gives
 $D_{\rm min}\sim 1/Q^2$ and the gap distribution
function
$\rho({\cal D})\sim {\cal D}^{-3/2}$. This result
has been obtained numerically in Ref.\cite{GKP91}.

\subsection{Bands. Stating a Problem}
\label{BWS}

The bandwidths are a more subtle matter.  Let us take $k$-th band belonging to
some generation with $P$-even at $\tau=-1$.  Then $E_k$ is the edge of this
band.  We can take one step back along the hierarchical tree.  We find
that the parent band $k'$ is of generation $\eta'=P'/Q'$ with $P'$-odd and
 $\tau=+1$.  The rational points at this generation give centers
of bands.  If we had started from the band $Q+1-k$ (symmetric to the $k$-th band)
we would find the band $Q'+1-k'$ as its parent band.  The roots of symmetric
bands of generation $\eta'$ are related by symmetry $z_{i}\rightarrow
-z_{i}$  (see Sec.\ \ref{appendix}).  Thus the string content of two paths leading
to the bottom and the top edges of the band differ only by the highest string
of the same length $l$ and the same parity $v_l$.  If we assume that adding
a string does not change the centers of already existing strings
(the noninteracting strings approximation), then the bandwidth is given just
by twice the energy of the highest string (\ref{energy03}).  This gives an
estimate of the bandwidth $W\sim \frac{1}{Q^2}$.  Corrections to this crude
estimate come from many sources (roots get some small corrections when the
new string is added, every root has ``finite size'' corrections) and these are
of the order $Q^{-2}$ as well.

It has been shown by Thouless \cite{ThoulessBW90} that the sum of all
bandwidths scales as $Q^{-1}$ (moreover $\lim_{Q\rightarrow\infty}Q\sum_k
W_k=const=9.32\ldots$).  If all bands scale similarly (fractal), then the
scaling dimension $Q^{-2}$ would fit the scaling of the total bandwidth.
We call this {\em the mean dimension}.  In reality, bandwidths scale differently
(multifractal).  Their anomalous dimensions $\epsilon_J$ depend on the path
$J$ of the hierarchical tree that the band belongs to: $W(J)\sim
Q^{-2+\epsilon_J}$, where $Q$ is the denominator of the flux running along
the path.  According to ref.\cite{HirKohm89} for $\eta=\frac{\sqrt 5-1}{2}$
exponents $\epsilon_J$ vary between $0.171$ and $-0.374$ , where the
positive anomalous dimension corresponds to a path in the center of the
tree, while the negative one belongs to the path along the lowest bands.

A distribution of anomalous dimensions of the bandwidths, can be described
by the function: $F(\beta)= -\lim_{Q\rightarrow\infty}\frac{\log
\sum_{k=1}^{Q} W_k^\beta}{\log Q}$ Here is a limited list of what  we do
know about this function.  The function $F(\beta)$ is convex and has linear
asymptotes $F(\beta)=\frac{\beta}{\epsilon_{\rm min}}$ for
$\beta\rightarrow -\infty$ and $F(\beta)=\frac{\beta}{\epsilon_{\rm max}}$
for $\beta\rightarrow\infty$.  A deviation of $F(\beta)$ from a straight
line (fractal) indicates multifractality.  
The point $\beta=D_{\rm H}$ 
where $F(D_{\rm H})= 0$ is known as the ``Hausdorff dimension'' and the 
slope of $F$ at this point, gives the most probable scaling dimension 
$\epsilon_{0} = F'(D_{\rm H})$. The scaling of the total bandwidth
\cite{ThoulessBW83,ThoulessBW90} gives $F(-1)=1$.  Obviously, $F(0)=-1$.
We have plotted $F(\beta)$ for the case where the flux is equal to the
golden mean in Fig.\ \ref{fig.fbeta}.

If $F(\beta)$ were a straight line, $D_{\rm H}$ would be $\frac{1}{2}$
and $\epsilon_{0}=2$. It would mean that each scaling dimension
$\epsilon (J)$ is equal to the mean field value 2.
 From Fig.\ \ref{fig.fbeta} we can see that $F(\beta)$ is indeed close to a
straight line. The Hausdorff dimension
$D_{\rm H}=0.49792\pm 4\times 10^{-5}$ and the most probable scaling dimension
$\epsilon_{0}=1.9996\pm 1\times 10^{-4}$
are numerically close to $\frac{1}{2}$ and $2$ respectively. Anomalous
dimensions are
small but code the most interesting features of the problem.

The topological classification of solutions to the Bethe equations by
means of strings alone gives the mean-field dimension of bandwidths and
absolute values of bandwidths with high accuracy of the order of $1/Q^2$.
Analytical determination of anomalous dimensions goes beyond just the
topological
  classification and is perhaps the most challenging problem in the field.
We do not attempt to solve it in this paper.

\section{Numerical Evidence for the String-Hierarchy Hypothesis}
\label{NumEv}

 In this Section we present a fragment of an extensive numerical
analysis we have performed in order to confirm the string-hierarchy
hypothesis of
the Sec.\ \ref{strings}. We choose the most ``irrational'' flux (golden
mean) $\eta=\frac{\sqrt{5}-1}{2}=\left[1;1,1,1,1\ldots\right]$. This is the
only
case where intermediate fractions are the same as convergents
\begin{eqnarray}
  \frac{P_i}{Q_i} & = &  \frac{F_{i-1}}{F_i}=\frac{1}{1},\frac{1}{2},
\frac{2}{3},\frac{3}{5},\frac{5}{8},\frac{8}{13},\ldots ,
\end{eqnarray}
where $F_i$ are Fibonacci numbers as well. They give the
fluxes
of  generations. The set of Hall conductances = Takahashi-Suzuki
numbers = allowed lengths of strings and are all Fibonacci numbers:
\begin{equation}
 2l_k+1 =  Q_{k-1} = F_{k-1} = 1,1,2,3,5,8,\ldots .
\label{gmtn}
\end{equation}

\begin{figure}
\centerline{\psfig{file=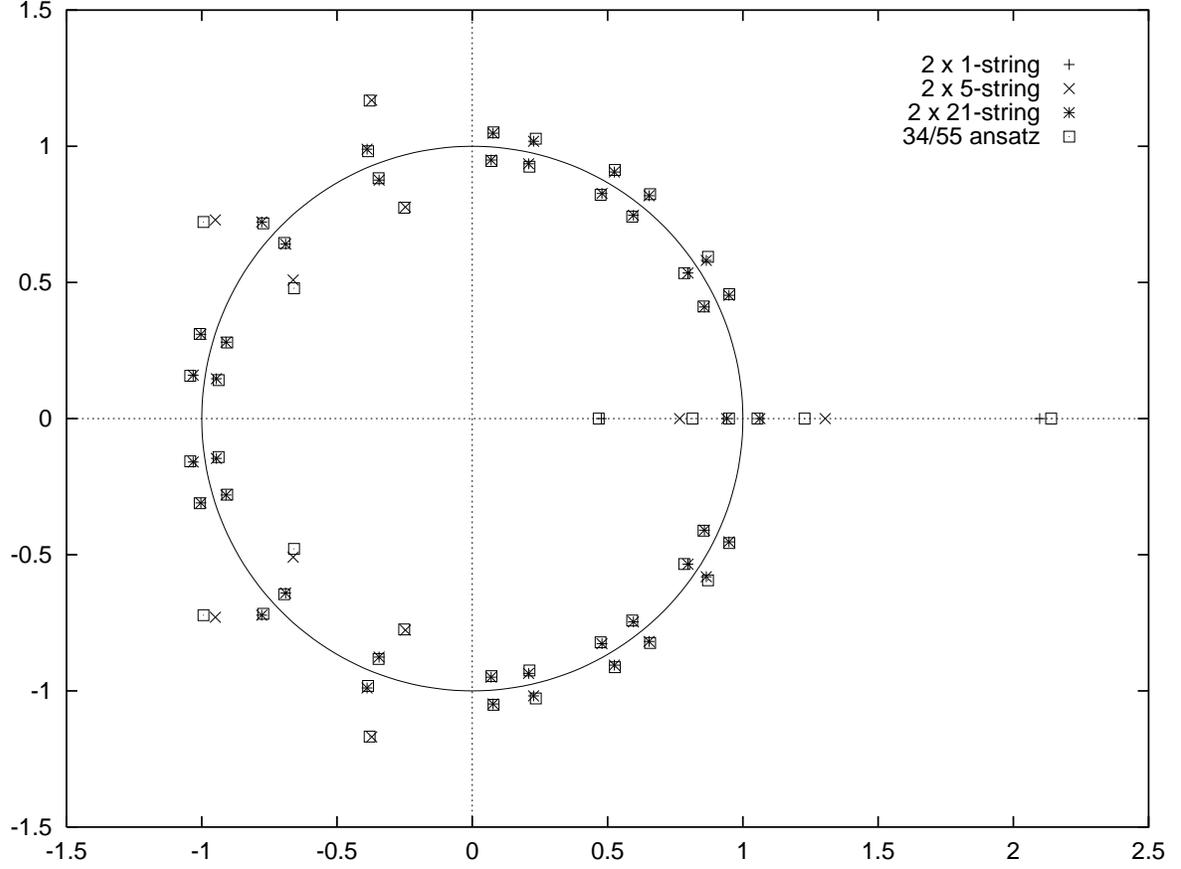,width=\hsize,angle=-90}}
\vspace{1.2cm}
\caption{Shown in the figure are the roots that parameterize the ground
 state of the Hofstadter problem at flux $34/55$. The roots are
 organized in strings: two of the length 1($+$), two of the
 length 5($\times$) and two of the length 21($*$). These roots can be
 approximated by ansatz (\protect\ref{emp-wf}) and are plotted in
 the Figure as boxes($\Box$). Strings or more generally roots tend to
 repel each other. This ``splitting'' is captured empirically by the
 expression for $x_n$ in (\protect\ref{emp-wf}).}
\label{fig.3455}
\end{figure}

As an illustration let us consider the bottom edge
of the lowest energy band of the generation
$\frac{34}{55}$. According to the
string hierarchy hypothesis the length of the largest string is $21$ 
and then iteratively one obtains that the string content consists of strings of lengths
$F_{3k+1} =1,5,21$ for $k=0,1,2$ and can be written as $(1,1,5,5,21,21)$. All
string parities in this example are $-1$. The actual numerically 
calculated roots are shown in Fig.\ \ref{fig.3455}. One notices 
immediately  that the roots organize themselves into distinct groups whose 
members have roughly the same radius: two groups with 1 root($+$), two with 
5 roots($\times$), and two groups with 21 roots($*$).  

The spins of strings and their parities are topological
characteristics of the state. In contrast to these properties, the centers  of
strings are not topological and become close to 1 only for large spin
strings. For our example they are
$x= 2.10$, $(2.10)^{-1}$, $1.23$, $(1.23)^{-1}$, $0.81$, $(0.81)^{-1}$,
$1.05$, $(1.05)^{-1}$. We have found numerically that the centers of
large strings are
$x_{F_{3n+1}}\approx 1+\frac{\xi}{F_{3n+1}}$ with $\xi=1.14$.

\begin{figure}
\centerline{\psfig{file=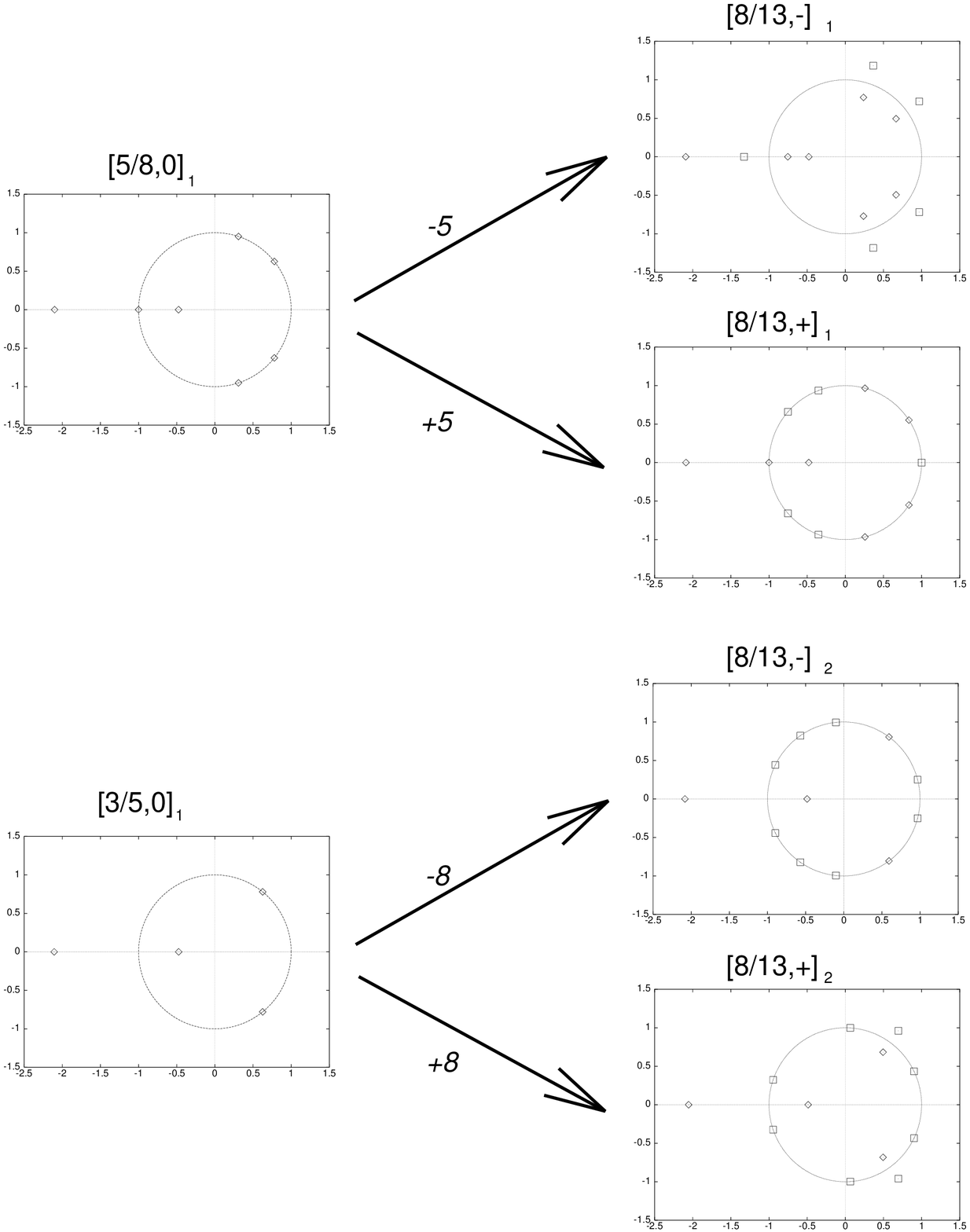,height=4in}}
\centerline{\psfig{file=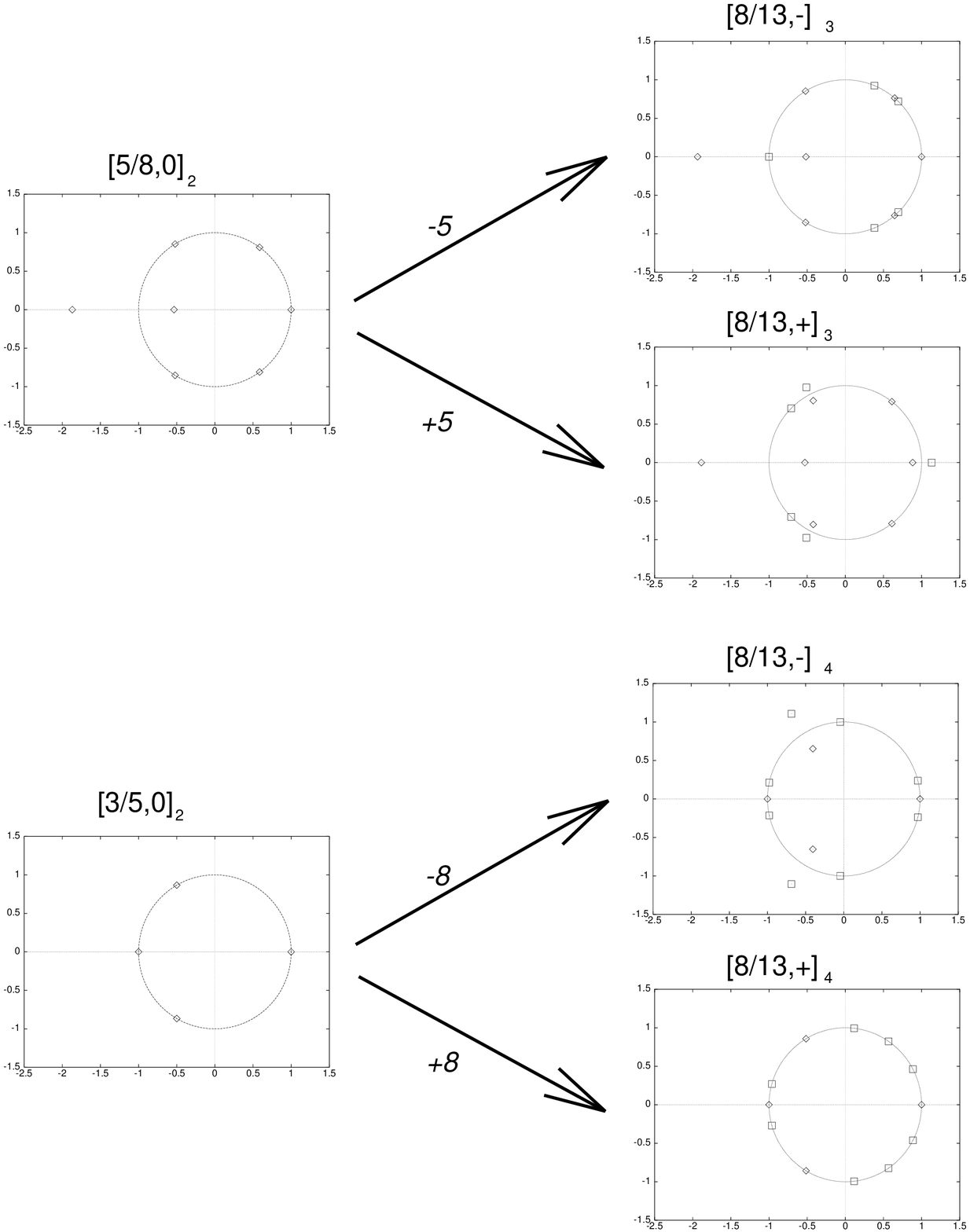,height=4in}}
\centerline{\psfig{file=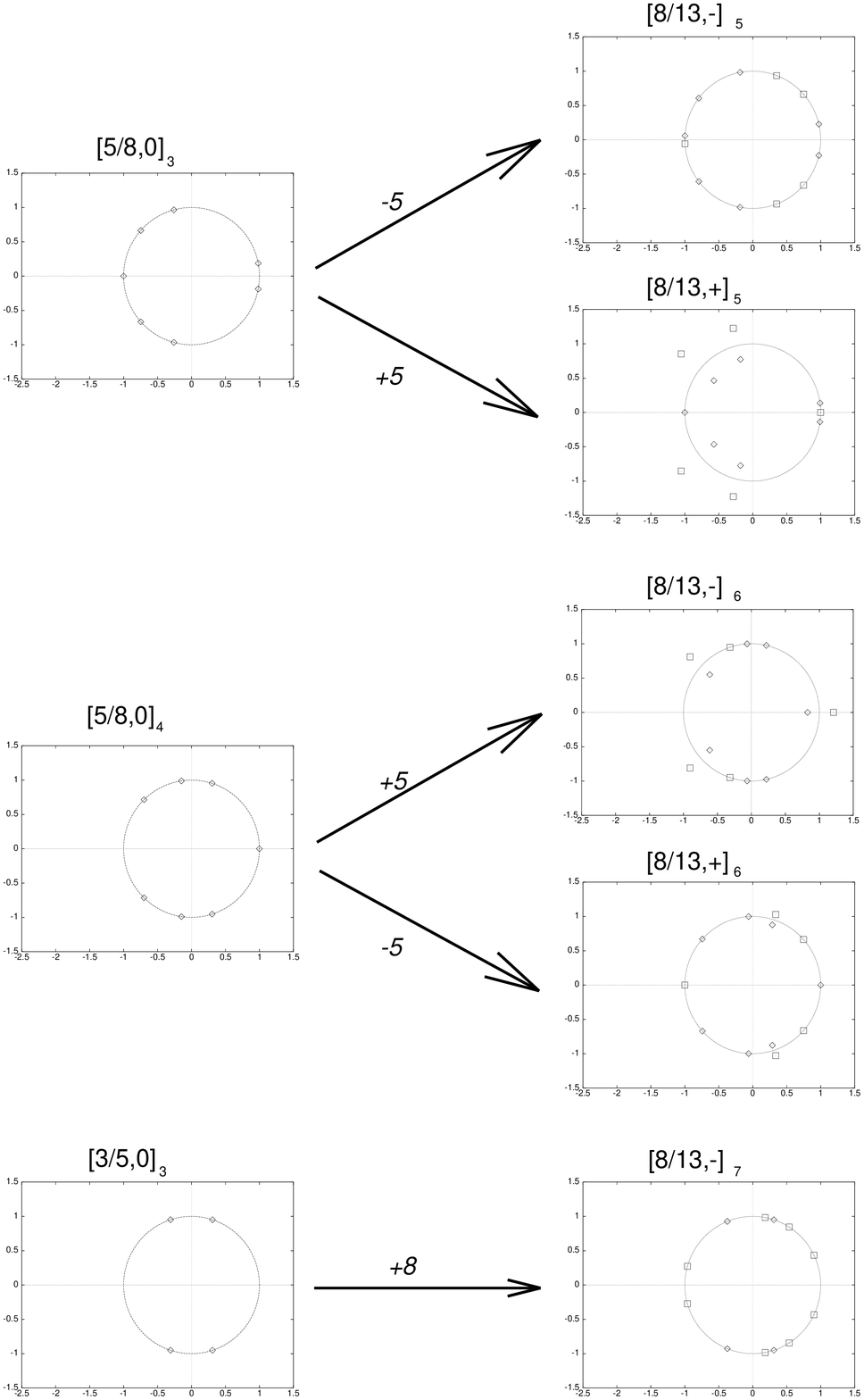,height=6in}}
\vspace{1.2cm}
\caption{This figure shows the evolution of root's pattern along the
hierarchical
tree: root patterns of the generation with the flux
$\frac{8}{13}$ are constructed out
  of previous generations with the flux $\frac{5}{8}$ and $\frac{3}{5}$) by
  adding strings with the lengths $5$ or $8$
  respectively with parities $\pm$ denoted as sign in front of string lengths. 
The symbol $\left[\frac{P}{Q},-0+\right]_i $
denotes a state  at the bottom  ($-$), middle
  ($0$) or top ($+$) of the $i$-th band (counted from the bottom of
the spectrum) of the generation with a
  flux $\frac{P}{Q}$ .}
\label{fig.examples}
\end{figure}

Another example, Fig.\ \ref{fig.examples}, illustrates the
hierarchy.
It shows how 12 roots
of the band {\em edges}
of  the generations $\frac{8}{13}$ can be approximately obtained
from 4 roots of the {\em centers} of the bands for the parent generation
$\frac{3}{5}$ and 7 roots of generation $\frac{5}{8}$($\Diamond$) by the
addition of
the strings of the lengths $8$ and $5$ ($\Box$) respectively.

The string decomposition gives an asymptotic form of the wave function
(\ref{PsiAnsatz}). In particular the wave function of the bottom edge of
the lowest
band of the generation $F_{3n-1}/F_{3n}$ is
\begin{eqnarray}
\label{emp-wf}
        \psi^{\{F_{3n-2},\ldots\}}(z) & \approx &
        \prod_{i=k}^{n-1}\prod_{m=1}^{F_{3i+1}}
        (z-x_{F_{3i+1}}q_{3i+1}^{-\frac{F_{3i+1}+1}{2}+m})
        (z-x_{F_{3i+1}}^{-1}q_{3i+1}^{-\frac{F_{3i+1}+1}{2}+m})
        \psi^{\{F_{3k-2},\ldots\}}
\end{eqnarray}
where $q_j=e^{i\pi \frac{F_{j-1}}{F_{j}}}$. This formula is
asymptotically exact at $k\rightarrow\infty$, but also works very well
even for $k=0,\;\psi^{\{0\}}=1$ and $\psi^{\{F_{3k-2},\ldots\}}$ is the
wave function of the generation $F_{3k-1}/F_{3k}$.

First of all let us check that the formula (\ref{emp-wf}) is
qualitatively correct and gives the same root pattern as the actual
roots of the Bethe Ansatz equations. In Fig.\ \ref{fig.3455} we
have 
presented, in addition to the numerically calculated roots corresponding to the ground state of
equation (\ref{FullFuncEq}) for the flux $\eta=34/55$, also the roots given by
(\ref{emp-wf}). It can be seen that the roots produced by
(\ref{emp-wf}) are in quite good agreement with the numerical
results. Actually one can check that the numerical accuracy with which
formula (\ref{emp-wf}) gives roots is $\sim 1/l^{2}$ for each root of the
string of the length $l$.
Now let us calculate the overlap between this wave function (\ref{emp-wf})
and the direct numerical solution of the eq.(\ref{HamChiral223}).

The result is tabulated in the following table, where the overlap
between the wave function obtained from the exact (numerically obtained)
roots $\psi_{\rm Exact}$ and ansatz wave function of different order
$k$: $\psi_{\rm Ansatz}^{\{k\}}$ is shown. The ansatz wave function of
the order $k$ is given by eq.(\ref{emp-wf}). It is ``made'' out of real
(numerical) and ansatz roots. The order $k$ of the ansatz function tells
how many of the longest strings are replaced by ansatz ones. $k=0$
corresponds to pure ansatz wave function. The bigger $k$ the more short
strings are replaced by ansatz strings. The positions of centers of the
ansatz strings are taken to be $x_{F_{3i+1}}=1+\xi/F_{3i+1}$ with the only
fitting parameter $\xi$ which is chosen to be $\xi = 1.14$.

\begin{center}
\begin{tabular}{|c|c|c|c|c|c|c|}
\hline
$k$ & $0$ & $1$ & $2$ & $3$ & $4$ & $5$ \\
\hline
$\langle \psi_{\rm Ansatz}^{\{k\}}|\psi_{\rm Exact}\rangle $ &
 $0.99475$ &
 $0.99464$ &
 $0.99657$ &
 $0.99669$ &
 $0.999435$ &
 $1.000000$ \\
\hline
\end{tabular}
\end{center}

The last column of this table reflects trivially the normalization of
wave function. One can notice that for $k=4$ when two strings of the
length $377$ are replaced by the ansatz roots the overlap between ansatz
wave function and exact one is $0.9994$ which is very close to 1.
Actually for this case even the purely ansatz wave function (with all
roots produced from (\ref{emp-wf})) has a good overlap: $0.995$ with the
exact one.

A more subtle check of the high degree of similarity between the exact
and ansatz wave function is the distribution of scaling exponents of
moments of the wave function, which is often used to characterize the
multifractal properties of the set generated by of values of $\psi_n $
\cite{HJKPS86}. This distribution is given by the Legendre transformation
\begin{equation}
\alpha =\frac{1}{F'(\beta)},\;\;
f(\alpha)=\beta-\frac{F(\beta)}{F'(\beta)}.
\label{FalphaDef}
\end{equation}
 of the function
\begin{equation}
F(\beta)=\lim_{Q\rightarrow\infty}\frac{\log\sum_{n=1}^{Q}
|\psi_n|^\beta}{\log Q}.
\label{FbetaDef1}
\end{equation}

An extended support of $f(\alpha)$ gives some measure of multifractality.

\begin{figure}
\centerline{\psfig{file=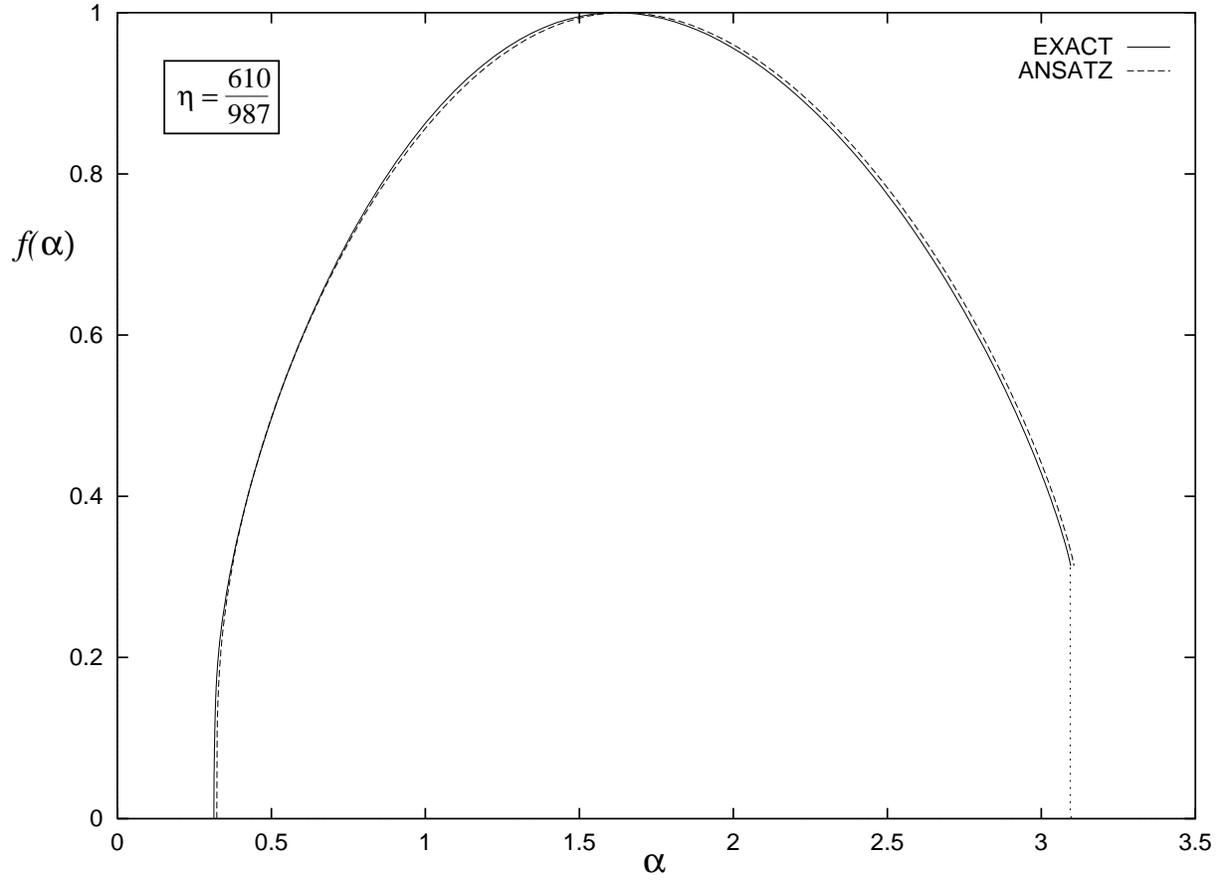,width=\hsize}}
\vspace{0.7cm}
\caption{The function $f(\alpha)$  for the  wave function of the 
bottom of the lowest band. The solid line is an exact result 
(found numerically),
the dashed line is the $f(\alpha)$ of the approximate wave function
(\protect\ref{emp-wf}). The support of $f(\alpha)$ is 
$0.3 \approx \alpha_{\rm min}\le \alpha \le \alpha_{\rm max} \approx 3.5$.}
\label{fig.falpha}
\end{figure}

\begin{figure}
\centerline{\psfig{file=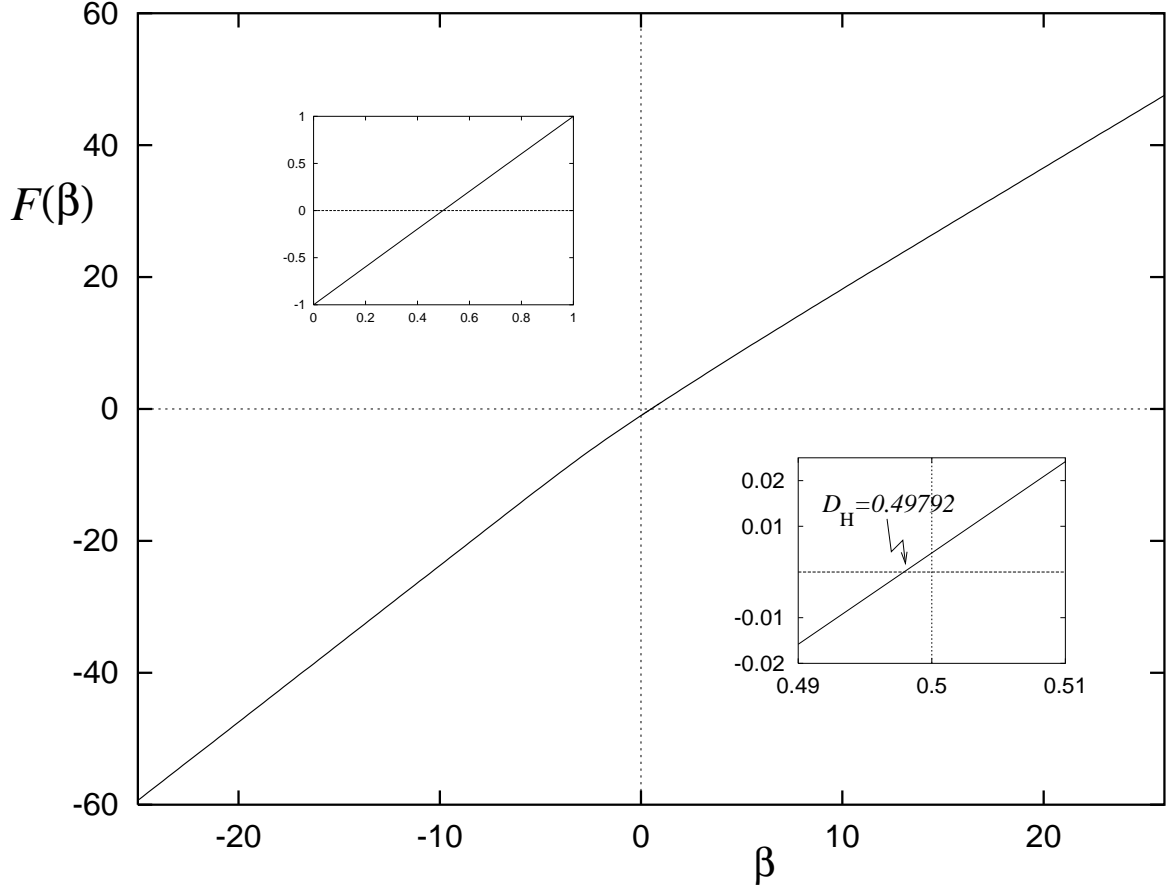,width=\hsize}}
\vspace{0.5cm}
\caption{The {\em free energy}
 $F(\beta)$ is plotted as a function of $\beta$ for the flux 
equal to the golden mean $\eta=\frac{\protect\sqrt{5}-1}{2}$.  The deviation from
a straight line indicates multifractality. The leftmost inset zooms into the region 
$\beta\in\left[0,1\right]$.  It includes the points
$F(\beta=0)=-1$ and $F(\beta=1)=1$  known  exactly. The rightmost inset zooms into the
region where $F(\beta =D_{\rm H})=0$. $D_{\rm H}$ is the Hausdorff
dimension. The slope of $F$ at this point  gives the most probable scaling
dimension $\alpha_0 = \frac{1}{F'(D_{\rm H})}$. Although both Hausdorff 
dimensions
and the most probable scaling dimension are numerically close to $0.5$ they
differ from $1/2$ within the numerical error-margin. The error margin
($10^{-8}$--$10^{-7}$) is not visible in the figure.}
\label{fig.fbeta}
\end{figure}

In Fig.\ \ref{fig.falpha} we have plotted $f(\alpha)$ for $\psi_{\rm
Ansatz}$ from eq.\ (\ref{emp-wf}) and $\psi_{\rm Exact}$ from the
exact roots, calculated numerically for a particular value of the flux, $\eta =
610/987$. By representing the data in this way we have neglected taking the
limit
in eq.\ (\ref{FbetaDef1}). For this value of the flux this is
justifiable by the fact that when increasing the denominator of the
flux beyond 987, $F(\beta)$ does not vary visibly. The figure shows that
the ansatz approaches the exact $f(\alpha)$ remarkably well, with just
one free parameter $\xi \approx 1.14$.

 Concerning the support of  $f(\alpha)$,
$\alpha_{\rm min}$ is approximately $0.32$ for the ansatz and $0.31$ for
the exact
wave function, whereas $\alpha_{\rm max}$ is approx.\ $3.5$ for both
ansatz and exact wave function within margin of error.

\section{Conclusions}
\label{conclusions}

The Azbel-Hofstadter problem as a typical quasiperiodic system  generates a
complex spectrum. Similar complex spectra were found in a number of dynamical
systems. Since the empirical observations of Hofstadter \cite{Hofstadter76},
evidence has been mounting that these spectra are regular and universal
rather than erratic or ``chaotic''. They have a deterministic hierarchical
structure.
In this paper we have shown that the topology of the strange sets,
generated in the problem is determined by the Chern numbers of the spectral
curve
i.e.\ by the Hall conductances. Even more so---at every finite step of the
hierarchy the spectrum is integrable. We have found an 
anticipated match between
Hall conductances and dimensions of representations of the quantum group
$U_q(sl_2)$. The latter are Takahashi-Suzuki numbers or the lengths of the
strings of the Bethe Ansatz solution. This correspondence suggests a natural
hierarchical tree, which, we believe, is relevant for general
quasiperiodic systems.

In the Bethe Ansatz language, each state is characterized by a particular string content.
Proceeding along the tree toward the incommensurate limit
corresponds to addition of strings. This picture is somewhat
reminiscent of the discrete renormalization group approach \cite{RGapproach}.

We were concentrated on the topological aspect of the string
hypothesis. It alone gives
\begin{itemize}
\item The explicit asymptotically exact form  of some wave functions for
irrational flux. Multi-fractal properties of these functions, although not exact,
are in good agreement with numerical results.
\item The scaling exponent $3/2$ of the gap distribution.
\item The most probable scaling dimension $1/2$ of the bandwidths.
\end{itemize}
However, the major ends of the strings are loose. The string hypothesis
solves the Bethe Ansatz equations with an accuracy ${\cal O}(Q^{-2})$, i.e.
is asymptotically exact in the incommensurate limit $Q\rightarrow\infty$. However,
the most interesting quantitative characteristics of the spectrum are actually in the
finite size corrections of the order of ${\cal O}(Q^{-2})$ to the bare value of
strings. Among them are the anomalous dimensions of the spectrum.  We
believe that it is possible to find them analytically via a more  detailed
study of
the Bethe Ansatz equations, beyond just the analysis of singularities. This
is a
technically involved but a fascinating problem. The ultimate solution of the
problem, however, would be through the application of conformal field theory approach, which has been
proven to be effective for finding the finite size corrections of 
integrable systems,
without the actual solution of the Bethe Ansatz.

\section{Acknowledgments}

We would like to thank J. Bellissard, who suggested  that the
noninteracting strings lead to the scaling exponent -3/2 for the gap
distribution
function and pointed out the Ref. \cite{GKP91}, and  Y.
Hatsugai for inspiring numerics during the initial stages of this project. We
acknowledge useful discussions with  Y. Avron, G. Huber, S.
Jitomirskaya, M. Kohmoto, Y. Last, R. Seiler and A. Zabrodin.
AGA was supported by MRSEC NSF Grant DMR 9400379. 
PBW was supported under NSF Grant DMR 9509533.

\section{Appendix: Bethe-Ansatz equations from Hofstadter model.}
\label{appendix}

In this section we (re)derive the Bethe Ansatz equations for the
Azbel-Hofstadter problem with an anisotropy parameter $\lambda$
\cite{FadKash95,Zabrodin94}.

The Hamiltonian of a particle on a lattice in a magnetic field  can be
written as
\begin{equation}%
\label{hfst}
H=T_x+T_x^{-1}+\lambda (T_y+T_y^{-1}),
\end{equation}%
where the translations operators on a square lattice in a magnetic field
$2\pi\eta$ per plaquette form a Weyl pair
\begin{equation}%
T_xT_y=q^2T_yT_x.
\end{equation}%
The representation $T_xf_n=f_{n+1},\;T_yf_n=q^{2n}e^{ik_y}f_{n}$
with $q=e^{i\pi\eta}$
gives the Harper equation (\ref{harper}). If the flux through a plaquette of
the square lattice is a rational fraction of the flux quantum $\eta =P/Q$ then the
operator $T_x^{Q}$ commutes with the Hamiltonian (\ref{hfst}) and there are finite
dimensional representations characterized by two momenta $k_x$, $k_y$ so that
\begin{equation}
T_x^Q=e^{iQk_x},\;T_y^Q=e^{iQk_y}.
\label{hfstbc}
\end{equation}
The non-equivalent representations are given by restricting $0\le k_x <2\pi/Q$ and
$0\le k_y <2\pi$ and Schr\"odinger equation becomes:
\begin{equation}
	E\psi_{n} = e^{ik_{x}}\psi_{n+1} +e^{-ik_{x}}\psi_{n-1}
	+2\lambda \cos(2\pi\eta n +k_{y})
 \label{origHarp}
\end{equation}
with periodic boundary conditions $\psi_{n+Q}=\psi_n$.
The Chambers relation \cite{Chambers65}
for the problem (\ref{origHarp})
says that the energy spectrum is obtained as the 
solutions of an algebraic equation $Pol(E)=\Lambda(k_x,k_y)$ where
$Pol(E)=E^Q+\ldots$ is a polynomial  of $Q$-th degree which coefficients do
not  depend on $k_x$ and $k_y$. The only dependence of energy on $k_x$
and $k_y$ comes from free term of algebraic equation via the
combination:
\begin{equation}
   \Lambda(k_x,k_y) =2\cos{Qk_x}+2\lambda^{Q}\cos{Qk_y}.
 \label{Lambda}
\end{equation}
Therefore the edges of the energy bands are given by extrema of $\Lambda$
which assumes a minimum/maximum given by $\Lambda=\pm 2(1+\lambda^Q)$.

Below we use another representation (the modified chiral gauge):
\begin{equation}
 \label{bc}
	T_x=UV\frac{U+a}{U+b},\;\;T_y=VU^{-1}\frac{U+a}{U+b},\;\;\;\;
	UV=qVU.
\end{equation}%
In this representation for the choice
\begin{eqnarray}
   a &=& -i\kappa \lambda^{-\frac{1}{2}} q^{-\frac{1}{2}}, \nonumber\\
   b &=& -i\tau\kappa \lambda^{\frac{1}{2}} q^{-\frac{1}{2}}, \nonumber
\end{eqnarray}
where $\tau,\kappa=\pm 1$
the Hamiltonian (\ref{hfst}) becomes
\begin{eqnarray}
H &=& q VU^{-1}(U-i\kappa \lambda^{-\frac{1}{2}} q^{-\frac{1}{2}})
(U+i\tau\kappa \lambda^{\frac{1}{2}} q^{-\frac{1}{2}})  \nonumber \\
  & &  + \lambda q^{-1}V^{-1}U^{-1}
(U+i\kappa \lambda^{-\frac{1}{2}} q^{\frac{1}{2}})
(U-i\tau\kappa \lambda^{\frac{1}{2}} q^{\frac{1}{2}}).
\label{hfstUV}
\end{eqnarray}
Let us choose the following representation of operators $U$ and $V$:
$$(U\psi)_n= e^{ip_-}q^{-n}\psi_n,\;\;
  (V\psi)_n=-i\tau\kappa e^{ip_+}\lambda^{\frac{1}{2}}\psi_{n+1}.$$
Then the Schr\"odinger equation corresponding to (\ref{hfstUV}) becomes
\begin{eqnarray}
   -\kappa q^{n}e^{-ip_{-}}\lambda^{-\frac{1}{2}}E\psi_n
     &=& iq e^{ip_+} (q^{n}e^{-ip_-}
   -i\tau\kappa\lambda^{-\frac{1}{2}}q^{-\frac{1}{2}})
   (q^{n} e^{-ip_-}
   +i\kappa\lambda^{\frac{1}{2}}q^{-\frac{1}{2}})
   \psi_{n+1} \nonumber \\
      &-&
   iq^{-1} e^{-ip_+} (q^{n}e^{-ip_-}
   +i\tau\kappa\lambda^{-\frac{1}{2}}q^{\frac{1}{2}})
   (q^{n} e^{-ip_-}
   -i\kappa\lambda^{\frac{1}{2}}q^{\frac{1}{2}})
   \psi_{n-1}
\label{HamChiral223}
\end{eqnarray}
The equation (\ref{HamChiral223}) has periodic coefficients and one
can require that $\psi_{n+Q}=\psi_n$ ($\psi_{n+2Q}=\psi_n$) for $P$-even
($P$-odd). Then the solutions will be labeled by $0 \le p_+ < 2\pi/Q$
for $P$--even and by $0 \le p_+ < \pi/Q$ for $P$--odd.

Let us now extend the periodic function $\psi_n
$ to the entire complex plane by writing
$\psi_n=\Psi(z)|_{z=-e^{-ip_-}q^n}$.
We obtain:
\begin{eqnarray}
   \kappa\lambda^{-\frac{1}{2}}zE\Psi(z)
  &=& iq e^{ip_+}(z
+i\tau\kappa\lambda^{-\frac{1}{2}}q^{-\frac{1}{2}})
(z-i\kappa\lambda^{\frac{1}{2}}q^{-\frac{1}{2}})
\Psi(qz) \nonumber \\
  &-&
iq^{-1} e^{-ip_+} (z
-i\tau\kappa\lambda^{-\frac{1}{2}}q^{\frac{1}{2}})
(z+i\kappa\lambda^{\frac{1}{2}}q^{\frac{1}{2}})
\Psi(q^{-1}z) .
\label{FuncEq}
\end{eqnarray}
If $e^{2ip_+}=1$,  the eq. (\ref{FuncEq}) has $Q$  polynomial
solution of the order $Q-1$. Below we denote
 $\mu = e^{ip_+}=\pm 1$. The change $\mu\rightarrow -\mu$ changes the 
sign of energy corresponding to given polynomial solution of (\ref{FuncEq}). 
Let us parameterize the polynomial solutions by its roots $z_i$:
\begin{equation}\label{jj}%
\Psi(z)=\prod_{i=1}^{Q-1}(z-z_i).
\end{equation}%

Evaluating eq.\ (\ref{FuncEq}) at one of the roots $z_i$,
we obtain the  Bethe Ansatz equations:
\begin{equation}
   q^Q\prod_{j=1}^{Q-1}\frac{qz_i-z_j}{z_i-qz_j} =
   \frac{\left(z_i-i\tau \kappa
   q^{\frac{1}{2}}\lambda^{-\frac{1}{2}}\right)
   \left(z_i
   +i\kappa q^{\frac{1}{2}}\lambda^{\frac{1}{2}}\right)}
   {\left(q^{\frac{1}{2}}z_i
   +i\tau\kappa\lambda^{-\frac{1}{2}}\right)
   \left(q^{\frac{1}{2}}z_i
   -i\kappa\lambda^{\frac{1}{2}}\right)}
 \label{BetheAnsatz}
\end{equation}
These equations have $Q$ independent solutions. Each of them consists of
$Q-1$ complex
numbers $z_k$, $k=1,\ldots,Q-1$. Comparing the coefficients of
$z^{Q}$ and $z$ on the left and right sides of equation (\ref{FuncEq}) one
finds the energy:
\begin{equation}
 \label{en1}
   E=i\mu  q^{Q}(q-q^{-1})\lambda^{1/2}
   \left[\kappa\sum_{i=1}^{Q-1}z_{i} -\frac{i}{q^{1/2}-q^{-1/2}}
   \left(\lambda^{1/2}
   -\tau \lambda^{-1/2}\right)\right]
\end{equation}
or
\begin{equation}
 \label{en2}
   E=i \mu (q-q^{-1})\lambda^{1/2}
   \left[\kappa\sum_{i=1}^{Q-1}-\tau z_{i}^{-1} -\frac{i}{q^{1/2}-q^{-1/2}}
   \left(\lambda^{1/2}
   -\tau \lambda^{-1/2}\right)\right].
\end{equation}
Combining these two formula we have a more convenient form of the energy
\begin{equation}
 \label{en3}
   E=i\mu \frac{1}{2} q^Q(q-q^{-1})\lambda^{1/2}
   \left[\kappa\sum_{i=1}^{Q-1}(z_i-q^{-Q}\tau z_{i}^{-1})
   -i\frac{1-q^{-Q}}{q^{1/2}-q^{-1/2}}
   \left(\lambda^{1/2}
   -\tau \lambda^{-1/2}\right)\right].
\end{equation}

The Bethe equations (\ref{BetheAnsatz}) possess some discrete symmetries: if the
set
$\{z_{i}\}$ is a solution with anisotropy
$\lambda$ and
energy $E$ then
$\{z_{i}^{*}\}$ is also a solution with the same energy and $\{-\tau
z_{i}^{-1}\}$
is a solution with the energy $q^{Q}E$. In its turn the set
        $\{z_{i}^{-1}\}$ is a solution of the problem with $\lambda^{-1}$, its
energy is $-\tau q^{Q}E$. Finally, there is a trivial symmetry of equations
$z_i,\kappa\rightarrow -z_i,-\kappa$ which does not change the energy.

After some algebra one finds a relation between the wave function of the Harper
equation  (\ref{origHarp})  and polynomial (\ref{jj}).

(i) In the case of $q^Q=-1$ ($P$-odd) and in the middle of bands the
relation is
\begin{equation}
   \psi_n = (-1)^n q^{-n^2+2n}\sum_{m=0}^{2Q-1}\prod_{j=0}^{m-1}
   \left(e^{ik_y} q^{2n+\frac{1}{2}}
   \frac{\lambda^{\frac{1}{2}}-\tau\kappa
         e^{-i\frac{k_x+k_y}{2}} q^{-j-\frac{1}{2}}}
        {\lambda^{\frac{1}{2}}-\kappa
         e^{i\frac{k_x+k_y}{2}} q^{j+\frac{1}{2}}}
   \right)
   \Psi(-i e^{i\frac{k_x+k_y}{2}} q^{m}) ,
\end{equation}
where $k_x,k_y$ satisfy the condition $\Lambda (k_x,k_y) =0$.

(ii) In the case of $q^Q=+1$ ($P$-even, $Q$-odd) and $\Lambda
=2\mu (-1)^{\frac{P}{2}}(\lambda^Q-\tau)$ the relation is
\begin{equation}
   \psi_n = q^{-n^2}\sum_{m=0}^{Q-1}\prod_{j=0}^{m-1}
   \left(e^{ik_y} q^{2n+\frac{1}{2}}
   \frac{\lambda^{\frac{1}{2}}-i\tau\kappa
         e^{-i\frac{k_x+k_y}{2}} q^{-j-\frac{1}{2}}}
        {\lambda^{\frac{1}{2}}+i\kappa
         e^{i\frac{k_x+k_y}{2}} q^{j+\frac{1}{2}}}
   \right)
   \Psi(- e^{i\frac{k_x+k_y}{2}} q^{m}) .
\end{equation}
If $\tau=+1$
it corresponds to the middle of bands and the rational points
 $k_x,k_y$  given by the equation
$\frac{\cos\frac{Q}{2}(k_x+\pi\frac{P}{2Q})}
{\sin\frac{Q}{2}(k_y+\pi\frac{P}{2Q})}
=\kappa(-1)^{\frac{Q-1}{2}}\lambda^{\frac{Q}{2}}$.
If $\tau =-1$, $\Lambda =2\mu (-1)^{\frac{P}{2}}(\lambda^Q+1)$ the relation
corresponds to  the points at the edges of bands
$k_x=\frac{\pi}{Q}\frac{1-(-1)^{\frac{P}{2}}}{2}$ and
$k_y=\frac{\pi}{Q}\left(\frac{1-(-1)^{\frac{P}{2}}}{2}+2l\right)$ for
$l=0,1,\ldots,Q-1$. If counted from the bottom of
the spectrum, these edges are ordered as bottom-top-bottom$\ldots$ if
$\mu(-1)^{\frac{P}{2}}=-1$  and top-bottom-top$\ldots$ if
$\mu(-1)^{\frac{P}{2}}=+1$.

\begin{thebibliography}{100}
\bibitem{tree}
A number of different hierarchical trees has been
discussed in Refs.\ \protect\cite{Hofstadter76},
\protect\cite{StinchcombeBell87}, \protect\cite{HAS89}.
\bibitem{ntnote}
For rigorous definition of best approximants and for correct
formulation of the theorem that all best approximants are given by
intermediate fractions see \protect\cite{Venkov70}.
\end{thebibliography}
\end{document}